\documentclass[fleqn,usenatbib]{mnras}

\usepackage{newtxtext,newtxmath}

\usepackage[T1]{fontenc}
\usepackage{ae,aecompl}
\usepackage{soul}

\usepackage{graphicx}	
\usepackage{amsmath}	
\usepackage{amssymb}	

\newcommand{\oone}{O~{\small I}}
\newcommand{\oonemath}{\text{O{\tiny I}}}
\newcommand{\ctwo}{C~{\small II}}
\newcommand{\ctwomath}{\text{C{\tiny II}}}
\newcommand{\sifour}{Si~{\small IV}}
\newcommand{\sifourmath}{\text{Si{\tiny IV}}}
\newcommand{\cfour}{C~{\small IV}}
\newcommand{\cfourmath}{\text{C{\tiny IV}}}
\newcommand{\cthree}{C~{\small III}]}
\newcommand{\cthreemath}{\text{C{\tiny III}}]}
\newcommand{\mgtwo}{Mg~{\small II}}
\newcommand{\mgtwomath}{\text{Mg{\tiny II}}}

\newcommand{\sithree}{Si~{\small III}]}

\newcommand{\althree}{Al~{\small III}]}

\title[High blueshifts at high-redshift]{New Constraints on Quasar Evolution: Broad Line Velocity Shifts over $1.5\lesssim z\lesssim 7.5 $}

\author[R. A. Meyer et al.]{
Romain A. Meyer,$^{1}$ \thanks{E-mail: r.meyer.17@ucl.ac.uk}
Sarah E. I. Bosman,$^{1}$
and Richard S. Ellis$^{1}$
\\
$^{1}$Department of Physics and Astronomy, University College London, Gower Street, London WC1E 6BT, UK\\}

\date{Accepted 2019 May 29. Received 2019 May 29; in original form 2019 February 11}

\pubyear{2018}

\begin{document}
\label{firstpage}
\pagerange{\pageref{firstpage}--\pageref{lastpage}}
\maketitle

\begin{abstract}
We present the results of a model-independent investigation of the rest-frame UV spectra from a comprehensive sample of $394$ quasars in the redshift range $1.5\leq z \leq 7.5$. We fit the main Broad Emission Lines (BELs) in the rest-frame range $1280 \text{ \AA} \leq \lambda \leq 3000 \text{ \AA}$ (\oone, \ctwo, \sifour, \cthree, \cfour\, and \mgtwo) with a lightly-supervised spline fitting technique. Redshifts are derived from the peaks of each fitted BEL and used to compute relative velocity shifts. We show that our method gives unbiased velocity shifts and is insensitive to spectral resolution and instrumental parameters. It is found that the average blueshift of the \cfour\, line with respect to several low-ionisation lines in luminosity-matched samples does not significantly evolve over $1.5\gtrsim z\gtrsim6$. However, the average blueshift increases significantly by a factor $\sim 2.5$ at $z\gtrsim 6$.
We propose that this redshift evolution can be explained by \cfour\, winds launched perpendicularly to an accretion disk with increased torus opacity at high-redshift, coupled with a potential orientation-driven selection bias. Our results open new exciting avenues of investigation into young quasars in the reionisation epoch.
\end{abstract}



\section{Introduction}

The number of known quasars in the early Universe has recently increased sharply, mainly due to the availability of deep wide-field surveys in the optical and the near-infrared. At the time of writing, we count more than $170$ quasars at $z>6$, and $5$ at $z>7$. This progress is due to the concerted efforts of different teams selecting and confirming candidates from diverse surveys such as the Sloan Digital Sky Survey \citep[SDSS,][]{Jiang2016}, Panoramic Survey Telescope and Rapid Response System data \citep[Pan-STARSS,][]{Kaiser2010,Banados2016}, the Dark Energy Survey - Visible and Infrared Survey Telescope for Astronomy (VISTA) Hemisphere Survey \citep[DES-VHS,][]{Reed2015}, the VISTA Kilo-Degree Infrared Galaxy Survey \citep[VIKING,][]{Venemans2013}, the VLT ATLAS survey \citep{Carnall2015}, the UKIRT Infrared Deep Sky Survey \citep[UKIDSS,][]{Venemans2007,Mortlock2009,Mortlock2011}, the DESI Legacy imaging Surveys (DELS) \citep{Dey2019}, the Wide-field Infrared Survey (WISE) survey, and, increasingly, from overlaps of some thereof \citep{Banados2018, Wang2018, Wang2018a,Yang2018, Fan2019}. The recent discovery of a gravitationally lensed quasar may suggest that we are missing further high-redshift quasars \citep{Fan2019,Pacucci2019}, and therefore it is possible their numbers will swell in the coming years.

This surge has enabled a large number of investigations into a variety of topics. Rising numbers of high-redshift quasars offer the possibility to study their central supermassive black holes. The high masses of the high-redshift black holes powering these quasars have proven a particular challenge to explain as they imply either high seed masses and/or continuous accretion at nearly the Eddington rate for most of their young life \citep[e.g.][]{Willott2010,Banados2018}. The study of the co-evolution of galaxies and quasars has also benefited from larger samples of younger quasars \citep[e.g.][]{Venemans2012,Decarli2017}. Determination of the quasar luminosity function at $z\gtrsim 5.5$ has constrained the contribution of quasars to the cosmic reionisation of hydrogen \citep[e.g.][]{Onoue2017,Kulkarni2018}.
Used as the most distant beacons of light in the Universe, quasars are also an invaluable tool to study the intervening material in absorption. Metal absorption studies have charted the cosmic densities of metals up to the first billion years \citep[e.g.][]{Becker2009,Ryan-Weber2009,DOdorico2013,Bosman2017,Codoreanu2018,Meyer2019}. The scatter in the Lyman-$\alpha$ forest at the tail end of reionisation has highlighted the patchy nature of this process \citep{Bosman2018,Eilers2018}. Finally, reduced absorption of the intrinsic Lyman-$\alpha$ emission by neutral gas in the so-called quasar `near-zone' has been explored as a promising test of the evolving neutral fraction of hydrogen at $z\gtrsim6$ \citep{Mortlock2011,Keating2015,Eilers2017,Banados2018}. 
\\ \newline \\
However, absorption studies rely on the accurate reconstruction of the quasar continuum. It is common practice to extrapolate the Lyman-$\alpha$ emission and the blueward continuum (both heavily absorbed in the reionisation era) from the redward continuum. The most common techniques are based on the fitting of spectral Principal Component Analysis (PCA) eigenvectors learnt from SDSS quasars \citep[e.g.][]{Suzuki2006,Lee2012,Paris2011,Davies2018}. Recent studies have debated whether or not these SDSS-PCA templates are suitable for early quasars \citep[][]{Mortlock2011,Bosman2015, Banados2018, Venemans2018, Davies2018}. A crucial question is hence whether the spectra of quasars evolve with redshift. On the one hand, absorption studies are based on the \textit{a priori} characterisation of the intrinsic continuum at low-redshift and the assumption that it does not evolve. On the other hand, the intrinsic evolution of the spectra of quasars would provide a key observation to constrain the evolution of Active Galactic Nuclei (AGN), their physics and their co-evolution with their host galaxies.

A prime candidate for testing the evolution of quasar spectra is the broad \cfour\, $\lambda 1549$ \AA \, emission line. The equivalent width of the \cfour\, line strongly anti-correlates with the continuum luminosity - the so-called Baldwin effect \citep{Baldwin1977}. The \cfour\, line is also known to present a variety of asymmetrical shapes and blueshifts, measured with respect to the quasar' systemic redshift, are found to be correlated with its equivalent width \citep[e.g.][and references therein]{Richards2011}. The origin of these peculiarities is often attributed to radiation-driven outflows or poorly understood Broad Line Region physics (BLR). The bulk population of low-redshift quasars \citep[e.g. SDSS,][]{Shen2008,Paris2017} present on average a mildly blueshifted \cfour\, line ($\sim 800$ km s$^{-1}$). Therefore, the extreme \cfour\, blueshifts ($\gtrsim 3000$ km s$^{-1}$) reported in the three first $z>7$ quasar \citep{Mortlock2011, Banados2018, Wang2018a} raised interesting questions about the possible evolution of quasar spectra (and the properties of the \cfour\, line in particular).

It remains to be seen, though, if the objects detected so far at high-redshift are representative of the whole population of early quasars. Recent observational studies of the \cfour\, line blueshift have provided contrasting evidence for such evolution \citep{Mazzucchelli2017,Shen2019}. 
Different methods for measuring the \cfour\, blueshift have complicated comparison across authors and redshifts \citep[e.g.][]{Coatman2016}. The determination of the systemic redshift against which \cfour\, velocity shifts are measured has variously been done \textit{via} the cross-correlation of a quasar template \citep{Hewett2010,Richards2011}, Independent Component Analysis (ICA) decomposition (Allen \& Hewett, in prep., used by e.g. \citealt{Coatman2016,Reed2019}), fitting PCA eigenvectors \citep[e.g.][]{Mortlock2011}, or using a reliable quasar Broad Emission Line (BEL) such as \mgtwo\, \citep[e.g.][]{Mazzucchelli2017,Shen2019,Reed2019}. Similarly, the location of the \cfour\, line has been variously determined via the line centroid \citep{Coatman2016,Reed2019}, the flux-weighted central wavelength \citep[e.g.][]{Mortlock2011} the peak of a multi-Gaussian (or Gauss-Hermite) profile \citep[e.g.][]{DeRosa2014,Shen2016,Shen2019} or the maximum of a PCA template \citep[e.g.][]{Paris2011, Paris2017}. Furthermore, quasar BELs can be broken down into broad and narrow components, which appear to shift differently \citep{Greig2017}, and absorption in the red or blue wing of the emission biases the measurement of the intrinsic shift. Some of these techniques may even hamper the opportunity to probe the possible evolution of quasar spectra as they explicitly rely on templates extracted solely from low-redshift objects. Such factors cast a shadow of uncertainty on these velocity shifts. 

In this paper, we aim to address the question afresh and study the evolution of quasar spectra across cosmic time. In order to achieve this, we have gathered a comprehensive set of quasars at $1.5<z<7.5$ based on various samples of optical and near-infrared spectroscopic observations. Crucially, we take advantage of \emph{QUICFit} \citep{Meyer2019}, a lightly user-supervised, model-independent spline fitting method for the quasar continuum and broad rest-frame UV lines redwards of Lyman-$\alpha$. This approach does not attempt to recover the systemic redshift, rather it simply retrieves in a uniform manner the observed peak of the BELs without assuming a particular model for these emissions.  We can thus study the relative shift between \cfour\, and \mgtwo, but also the relative shifts between \oone, \ctwo, \cthree, \sifour \, and the two previous lines as other potential markers of evolution. In doing so, we can probe the shifts of high-ionisation lines with respect to low-ionisation lines across cosmic time to evaluate on possible quasar evolution and provide insight into BLR physics. 

The outline of this paper is as follows. In Section \ref{sec:methods}, we first detail our different datasets and the selection procedure of luminosity-matched control sample across our redshift range. We then assess the quality of our fits, and the different biases that could affect our measurement. Finally, we compare our method to PCA and Gaussian fitting on SDSS quasars. We present in Section \ref{sec:results} the evolution of the relative velocity shifts of all aforementioned BELs with respect to one another. In Section \ref{sec:discussion}, we discuss our results in light of previous studies and we propose a novel interpretation of the increased \cfour\, blueshift observed in early quasars.

Throughout this paper, we adopt a $\Lambda$CDM cosmology with $\Omega_M=0.3$, $\Omega_{\Lambda}=0.7$, $H_0 = 70 \text{ km s}^{-1}\text{Mpc}^{-1}$. We adopt a positive sign for velocity redshifts and negative for blueshifts.

\section{Methods}
\label{sec:methods}

\subsection{Observational samples}
\label{sec:methods_samples}

\begin{figure}
    \centering
    \includegraphics[width = 0.5\textwidth]{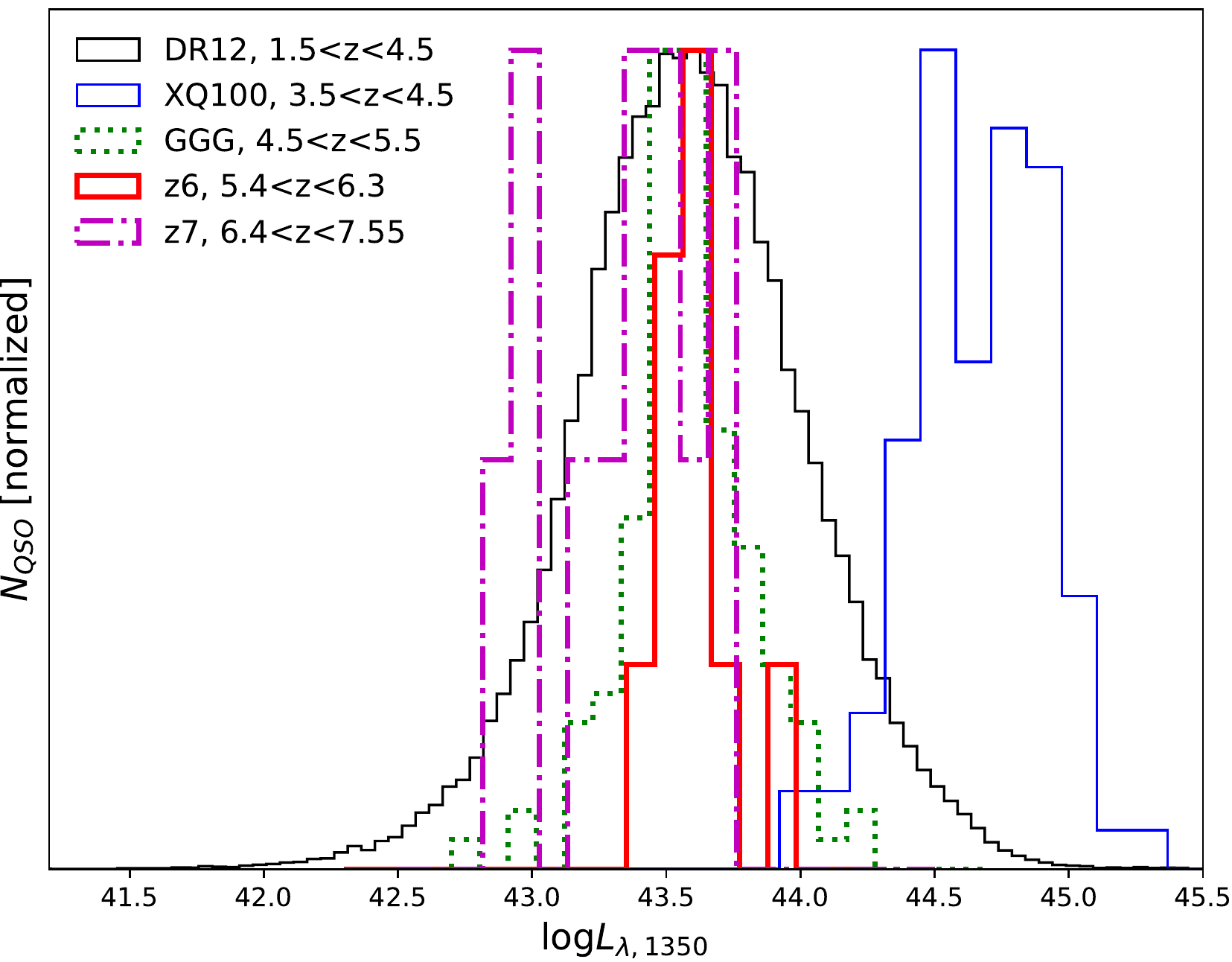}
    \caption{Specific luminosities (derived from the observed specific flux at $1350$ \AA\, $f_{\lambda, 1350}$) of our different samples. Note the good match between the luminosities of the GGG (green), z6 (red) and z7 (magenta) samples. They can then be readily compared to each other and to luminosity-matched subsamples of SDSS DR12 (black). XQ100 (blue), by contrast, contains significantly brighter objects. }
    \label{fig:samples_luminosities}
\end{figure}

\begin{figure*}
    \centering
    \includegraphics[width = \textwidth]{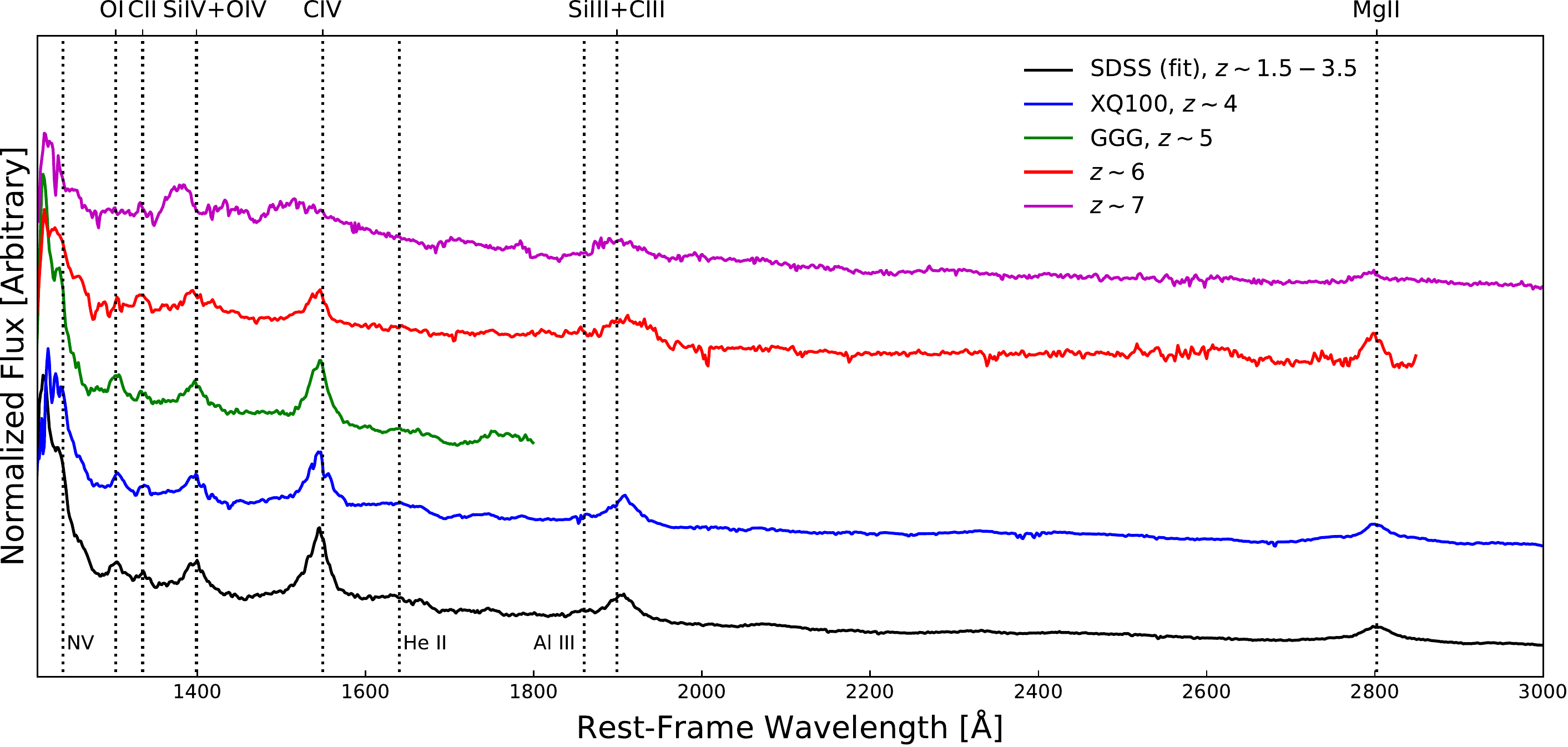}
    \caption{Rest-frame UV mean spectra of our different samples, presented with increasing redshift from bottom to top (black: SDSS luminosity-matched fitted subsample, blue: XQ100 (significantly higher luminosity than the other samples presented here), green: GGG, red: z6 quasars, magenta: z7 quasars). The spectra are normalized by their flux at $1350$ \AA\, and the powerlaw is not removed before stacking. We indicate the six major BELs we fit with vertical dashed lines. The \cfour\, emission can be seen to blueshift and to present a lower equivalent width in the high-redshift quasars. The mean \cfour\, line of the $z\sim 7$ stacked spectra is extremely flat due to the huge spread of blueshifts between a relatively small number of objects, but the mean shift is still visible. Other noticeable features include the reduced Lyman-$\alpha$ emission at high-redshift due to an increased neutral fraction of hydrogen and the emergence of \sifour\, at the expense of O{~\small IV} over $5\lesssim z\lesssim 7$.  }
    \label{fig:mean_spectra}
\end{figure*}

In order to study the velocity shifts of quasar UV BELs across cosmic time, we have gathered a large sample of quasar spectra in infrared and optical light from different surveys and archival data. We briefly describe here the original samples from which our quasars were drawn and how we construct our luminosity-matched control samples. The relevant parameters of the studied datasets are summarized in Table \ref{tab:samples} and the luminosity distributions are presented on Figure \ref{fig:samples_luminosities}. We also show the stacked spectra of all our samples on Figure \ref{fig:mean_spectra}. Throughout this study, we refer to the specific luminosity at $1350$ \AA, which we compute from the specific flux at $1350$  \AA\, rest-frame and the luminosity distance, simply as the \textit{luminosity}. 
We note that although the $1350$ \AA\, luminosity is a practical choice given it can be measured for nearly all of our quasars, a different choice of luminosity matching (e.g. rest-frame X-ray or IR) might affect the results of this study due to dust correction effects.
However, the emissivity at $1350$ \AA\, is more closely related to the UV emission lines of interest. In addition, the overlapping spectral coverage for our large sample of quasars is limited to the rest-frame UV. 

\begin{table}
    \centering
    \begin{tabular}{lllllll}
Name & $\Delta z$ & $\Delta L_{\lambda,1350}$ & R & $\overline{ \text{SNR}}$ & $N$ \\ \hline
DR12Q$^{a}$ & $1.5-2.4$ & 40.6-46.0  & 1500-2500 & 3.4 & 99232  \\
SDSS$^{b}$ & $1.5-4.5$ & 43.3-44.4 &1500-2500 & 14 & 108 \\
XQ100 & $3.5-4.5$ & 44.0-45.2 & 6000-9000 & 30 & 100  \\
GGG & $4.4-5.5$ & 42.7-44.3 &  800-900 & 20 & 163  \\
z6 & $5.4-6.3$ & 43.4-44.3 & 9000-11000& 90& 11 \\
z7 & $6.4-7.5$ & 42.8-43.8 & $\sim 6000$ & 6 &   12 \\
    \end{tabular}
    \caption{The different quasar samples with their redshift and luminosity ranges, resolution, median SNR and number of objects. (a) We use here the subset of DR12Q that have both tabulated \cfour\, and \mgtwo-derived redshift. (b) Throughout the text, SDSS refers to a subsample of the DR7Q and DR12Q overlapping quasars, luminosity-matched to the z6 sample, described in Section \ref{sec:methods_samples}}
    \label{tab:samples}
\end{table}

\textbf{SDSS DR7 and DR12:} SDSS-I/II, -III \citep{York2000,Eisenstein2011} delivered nearly $5$ million spectra and $32000$ deg$^{^2}$ of imaging in the \textit{ugriz} filters. In particular, we are interested here in two major catalogues of quasars observed and identified within SDSS data: the DR7 quasar catalogue \citep[DR7Q,][]{Schneider2010} and the DR12 quasar catalogue \citep{Paris2017}. For DR7Q, we make use of the value-added catalogue of \citet{Shen2011} which provides additional properties for the quasars listed in the original DR7 catalogue. \citet{Shen2011} report redshifts derived from \cfour, \cthree\, and \mgtwo, and the specific luminosities. In DR7Q, the line-based redshifts (i.e. redshifts derived from one emission line) are derived by fitting a multi-Gaussian template to each broad emission line and taking the template peak to derive the redshift. In DR12Q however, all quasars were inspected by eye and have thus a visual redshift, which is different from the line-based redshifts. The line-based redshifts in DR12 are derived from the peak of a PCA template fitted locally for each emission line \citep{Paris2011}. There is some overlap between DR7Q and DR12Q, although the latter sample is larger and contains quasars with slightly lower luminosities and in a broader redshift range. The SDSS optical spectrograph \citep{Smee2012} covers the range $3600-1000$ \AA, with a resolution going from $R\sim 1500$ to $R\sim 2500$. Hence the detection of both \cfour\, and \mgtwo\, is limited to quasars at $1.5 \lesssim z \lesssim 2.4$. However, other UV BELs like \oone, \ctwo\, or \sifour\,  can be detected simultaneously with \cfour\, in the complementary and overlapping redshift range $1.8\lesssim z \lesssim 5.1$. 

\textbf{XQ100:} XQ100 \citep{Lopez2016} is a legacy European Southern Observatory (ESO) survey which observed 100 quasars at $3.5 \leq z \leq 4.5$ at medium resolution ($R\sim 6000-9000 $) and a high median Signal-to-Noise Ratio $SNR \simeq 30$ with the XShooter instrument \citep{Vernet2011} at the Very Large Telescope (VLT). Its primary aims included absorption lines studies, AGN science, Broad Absorption Lines (BAL) studies, Damped Lyman-$\alpha$ absorbers (DLAs) and Lyman-$\alpha$ forest-based cosmology.  The combined wavelength range of XShooter's UVB, VIS and NIR arms ($3150-24800$ \AA, or $3150-18000$ \AA, depending on at which point in the survey the object was observed) is well suited to the aim of this study as all UV BELs can be retrieved over the entire redshift range $3.5 \leq z \leq 4.5$, except when they fall in telluric regions, spectrograph arm overlaps and parts impeded by the atmospheric transmission. XQ100 also overlaps partially with SDSS, making this subset ideal to study the impact of resolution on the observed spectra and the recovered fitted continuum. However, the XQ100 quasars were selected to be intrinsically bright, making them difficult to luminosity-match with some of the high-redshift samples. 

\textbf{Giant Gemini GMOS Survey:} The Giant Gemini GMOS survey  \citep[GGG, ][]{Worseck2014} obtained spectra of quasars over $4.4\leq z\leq 5.5$ with the Gemini Multi Object Spectrometers (GMOS) to study the Lyman-continuum (LyC) flux at $z\sim 5$ and determine the LyC photon mean free path. The publicly available data release of the GGG survey is composed of $163$ quasars at low-resolution ($R\sim 800-900$) with a medium-high $\text{SNR} \sim 20$, making it also suitable to capture the broad shapes of the UV BELs. The GMOS instrument covers the range $ 4800 \AA \lesssim \lambda \lesssim 10200 \AA$, meaning that \cthree \, and \mgtwo\, can only be observed for quasars at $z\leq4.4$ and $z\leq 2.6$, respectively. Thus we cannot measure the \cfour-\mgtwo\, relative shifts. However, since we do not restrict ourselves to this particular pair of BELs, we can use GGG quasars to compute velocity shifts of lower rest-frame wavelength BELs. GGG thus suitably completes the redshift ladder we have built by filling the gap between the SDSS datasets, XQ100 quasars and the highest-redshift samples at $z>5.5$.

\textbf{$z\sim 6$ sample: } Our first high-redshift sample (hereafter `z6') consists of a subset of quasars from the extensive list of $z\sim 6$ object from \citet{Bosman2018}, of which some where re-reduced later by \citet{Meyer2019}. We keep only those with near-infrared coverage and high SNR, limiting the sample to $11$ objects. This approach ensures a comparison as complete as possible over $1.5 \leq z \leq 6.5$, including all the lines from \oone\, to \mgtwo. This sample is critical to probe with additional spectra claims in the literature of (non-)evolution from the \cfour-\mgtwo\, velocity shifts. The $11$ quasars selected have both the highest-resolution ($R\sim 9000-11000$) and highest median SNR ($90$) of our samples (see Table \ref{tab:samples}).

\textbf{$z\sim 7$ sample: } Our second high-redshift sample is composed of the \citet{Mazzucchelli2017} quasars that are observed far enough in the infrared to detect at least \cfour, to which we add the present record-holding $z=7.54$ quasar \citep{Banados2018}, kindly provided by private communication of the authors. This sample thus comprises $12$ quasars observed with various instruments with $\langle z \rangle = 6.73$, median $\text{SNR} \sim 6$ (although individual spectra range from $\sim 2$ to $\sim 100$) and resolution $R\sim 6000$. 

\label{sec:subsdss}
Since BEL shifts correlate with luminosity, studying their potential evolution with redshift requires doing so at a fixed luminosity. We therefore construct luminosity-matched control samples from the SDSS quasars for all our higher-redshifts samples by oversampling their luminosity distribution. For for each quasar in a test high-redshift sample (XQ100, GGG, z6, z7), we draw a random SDSS quasar with a similar luminosity ($\Delta L = 0.2$ dex). We then repeat the procedure until we reach $5000$ SDSS quasars. DR7Q and DR12Q only contain information about the peak location of \cthree, \cfour\, and \mgtwo, derived with different methods highlighted above which cannot be compared directly with our measurements. To enable a fair and uniform comparison, we also construct a small luminosity-matched (to z6) control sample of $108$ quasars spanning the redshift range $1.5<z<4.5$. This smaller control sample can be then fitted with our spline technique to control our systematics between different peak retrieval methods, resolution and instruments, and derive \oone, \ctwo\, and \sifour -based redshift for SDSS quasars as well.

\subsection{Continuum fitting and line-derived redshifts}
We now describe the continuum fitting method and the subsequent derivation of the line-based redshifts that we apply uniformly to the quasar samples described above.

To enable the algorithm to run on a globally flat spectrum, each quasar is fitted with a continuum powerlaw \citep[e.g.][]{VandenBerk2001} with the following form
\begin{equation}
    F(\lambda) = F_0 \left(\frac{\lambda}{2500 \AA} \right)^\alpha
\end{equation}
where $F_0$ [$10^{-16}$ erg s$^{-1}$ cm$^{-2}$ Hz$^{-1}$] is the specific flux at $2500 $\AA, $\lambda$ the rest-frame wavelength in \AA\, and $\alpha$ is the powerlaw exponent. The powerlaw is fitted to regions relatively devoid of features following \citet{Decarli2010,Mazzucchelli2017} at  $1285-1295, 1315-1325, 1340-1375, 1425-1470, 1680-1710, 1975-2050, 2150-2250$ and $2950-2990$ \AA \, rest-frame with a positive linear prior on $F_0$ and $\alpha$.

Once each rest-frame UV spectrum spectrum is divided by the powerlaw, we fit it using \textsc{QUICFit}, which we describe briefly here. Further details can be found in \citet{Meyer2019}. \textsc{QUICFit} is a lightly-supervised spline quasar continuum algorithm which aims to avoid the human selection of appropriate continuum portions of the quasar spectra.
Instead, the algorithm first discards BALs (typically of equivalent width EW$>1-2$ \AA) by running a Gaussian matched-filter with a width of a few \AA ngstr\"oms. Smaller emission and metal absorption features, as well as bad pixels and cosmic rays are removed by searching for portions of the spectrum with excessive variance in the pixel-to-pixel flux differences. Indeed, if the spectral resolution is much higher than the continuum typical variation scale, the pixel-to-pixel flux increments and decrements distribution should follow the error array distribution. By computing the empirical variance in small portions of the spectrum, \textsc{QUICFit} can efficiently detect and removes structures (such as narrow absorbers, emitters or cosmic rays) that differ from a smooth continuum and expected uncertainty. Once this is done, the remaining parts of the spectrum can be considered representative of the intrinsic continuum and used to fit third order splines. We subsequently call these selected parts of the spectrum `continuum pixels' later in the paper. 

The spline knot points (where the polynomials of the spline join and the continuity conditions are enforced) can be initialized by the user, but it is found that one knot point at the location of each BEL, and one intermediate point in between, are sufficient to fit adequately third-order splines on the pre-selected continuum pixels. The process is lightly supervised and fast (a minute or two per quasar), as the user only has to adjust a threshold to reject pixels in narrow absorptions based on the expected uncertainty (see above) and choose whether or not to fit faint BELs by adding more spline knots. It is also found that complex lines such as \cfour\, or \cthree\, are better fitted with an extra knot point. Therefore we take particular care in accurately fitting  the emission lines of interest: \oone, \ctwo, \sifour, \cthree, \cfour\, and \mgtwo. This approach avoids any misfits occurring in pipelines for large datasets such as SDSS quasars, while being still applicable to smaller datasets with a range of SNR and resolution.

\begin{figure*}
    \centering
    \includegraphics[width=\textwidth]{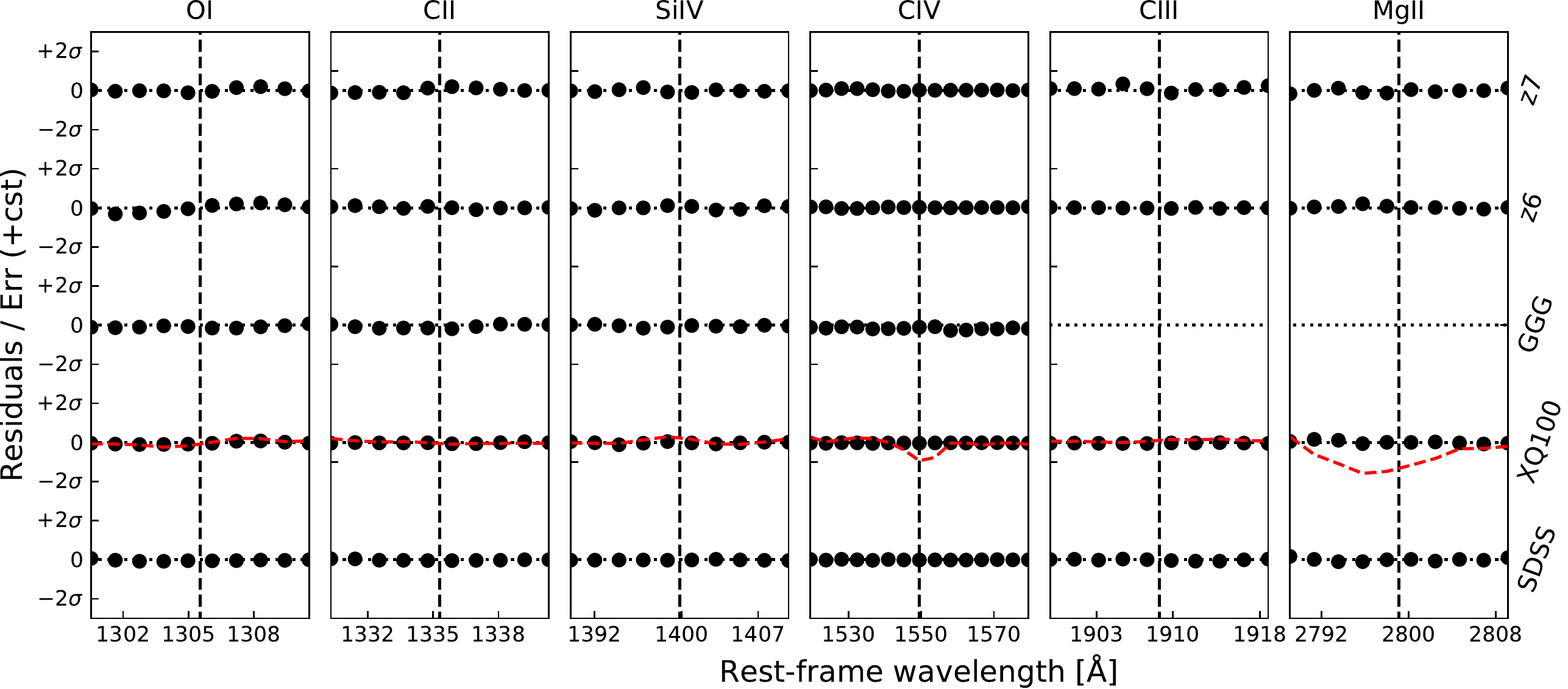}
    \caption{Stacked and binned residuals of the fitted splines (black points) for the six different UV BELs of interest in our five different samples. The residuals show a very good agreement between the continuum-selected pixels and the resulting spline fit. The SDSS sample is a $108$ quasar-strong sub-sample of SDSS chosen to match the luminosity of the z6 sample (see Sec. \ref{sec:subsdss}). The red dashed line shows the residuals between the high-resolution XQ100 spectra and the continuum splines recovered from a degraded resolution version of the same spectra. The deviations at the $1-1.5\sigma$ for \cfour\, and \mgtwo\, are expected from the widening of narrow absorbers damping the observed peak at low resolution.}
    \label{fig:stacked_residuals}
\end{figure*}

The resulting spline fits and the line-based redshift solutions for all the BELs of our quasars are presented in Appendix \ref{appendix:fits}. Before investigating the suitability of our scheme to recover unbiased measurements of the BELs peak, we first focus on the quality of the fits by looking at the residuals. In order to do so, we assume that our selection of continuum pixels is correct, and compute the stacked residuals on these points only. We present on Fig. \ref{fig:stacked_residuals} (black points) the residuals of the continuum fits for all the different samples. The residuals are flat and show no particular trend, evidencing that we fit correctly the continuum of the selected continuum pixels. 

Once the BELs have been fitted, we retrieve the peak of the spline fit to derive the line-based redshifts. In doing so, we take care to select only significant relative maxim in the continuum according to the following criteria. We first take the maximum in a rest-frame window of $10$ \AA \,(\oone, \ctwo, \sifour, \cthree) or $40$ \AA \, (\cfour, \mgtwo) around the expected BEL location, computed using the reported redshift of the quasar in the literature.  We then require that this maximum is significantly higher than the surrounding continuum. In order to do so, we take the relative minima of the continuum at lower and higher wavelength and we interpolate the continuum between the two minima. We compute the equivalent width (EW) of the line with respect to the interpolated continuum and we retain the line if it has $\text{EW}>3$\AA. As the EW of the \cfour\, line anti-correlates with its blueshift \citep[e.g.][]{Coatman2016}, we expect such a threshold could potentially bias the measurement of the blueshifts towards lower values. However this should not affect a potential evolution of the average blueshift. 
If multiple such maxima are retained, as can be the case for the \sifour\, and \cthree\, blended multi-line complexes, we retain only the lower and higher wavelength peak, respectively. This is because we are interested in the \sifour\, and \cthree\, emissions, not the blended O{\small ~IV} $\lambda 1402$ \AA\, and \sithree\,$\lambda 1892$ \AA, respectively. This procedure disentangles the peaks when the two emission lines have similar equivalent widths and are resolved. Nonetheless, in some cases the line-based redshift simply tracks the peak of the blend, and is thus slightly biased towards larger (lower) values for \sifour\, (\cthree). We further comment on the impact on our results and the suitability of these line complexes to study relative velocity shifts in Section \ref{sec:results}.  

\subsection{Instrument and resolution biases}
\label{sec:methods_res}.

Our quasar samples (Table \ref{tab:samples}) have different resolutions and SNR, and have been observed at different redshifts. We have fitted the continuum with QUICFit in a uniform way for the SDSS, XQ100, GGG, z6 and z7 samples. However, we need to ensure that our method is not biased by the varying resolution of the different samples towards lower or higher velocity shifts. This might be the case in the presence of a large number of unresolved associated narrow absorbers that would blend with the intrinsic profile and damp the blue or the red wing of the line. At higher resolution, these absorbers are resolved and thus successfully removed from the continuum to recover the intrinsic profile. This might create a marked difference between the high- and low- resolution samples if absorbers are found preferentially in the blue or red wing of BELs.

\begin{figure*}
    \centering
    \includegraphics[width=\textwidth]{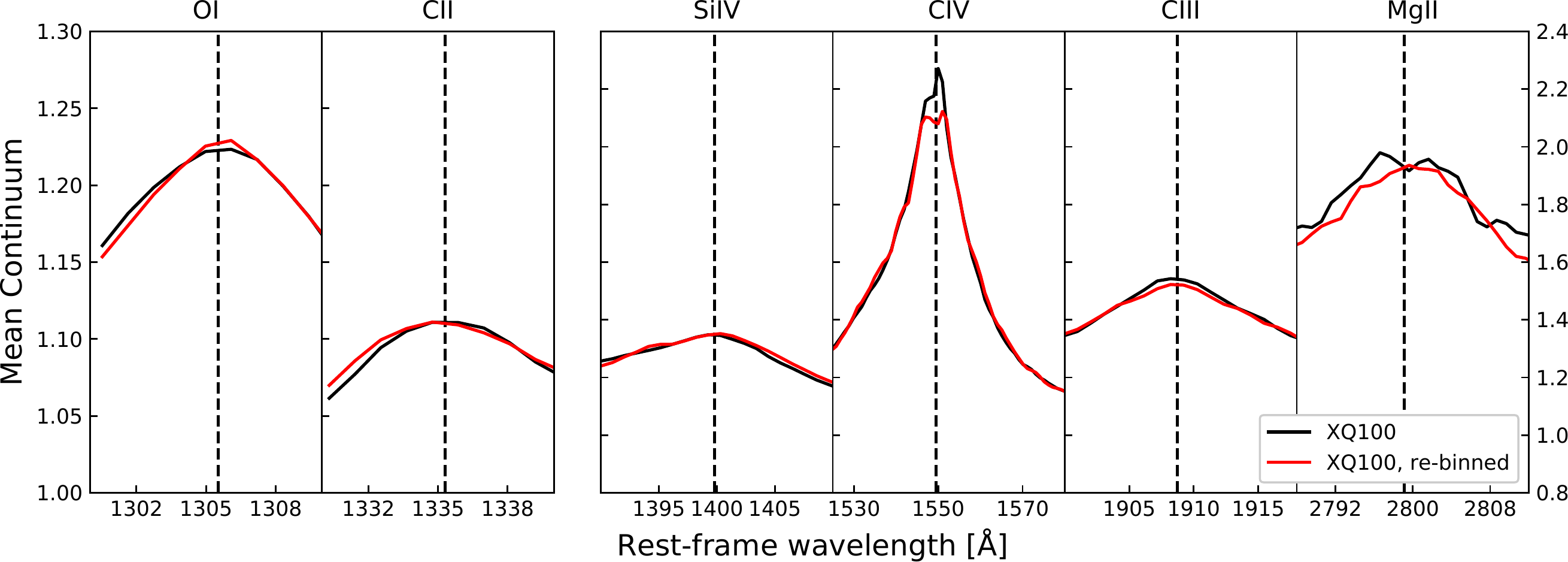} 
    \caption{Comparison of the stacked emission profile for the six BELs of interest obtained from the regular ($R\sim 7500$,black) and degraded resolution ($R\sim 1800$,red) version for the XQ100 quasars. While the \oone, \ctwo, \sifour\, and \cthree\, continuum are essentially identically recovered, \cfour\, and \mgtwo\, are impacted by the widening of narrow absorbers at lower resolution. The emission profile recovery with \emph{QUICFit} is thus quite robust to a change in resolution.  \label{fig:continuum_res_change}}
\end{figure*}

\begin{figure*}
    \centering
    \includegraphics[width=\textwidth]{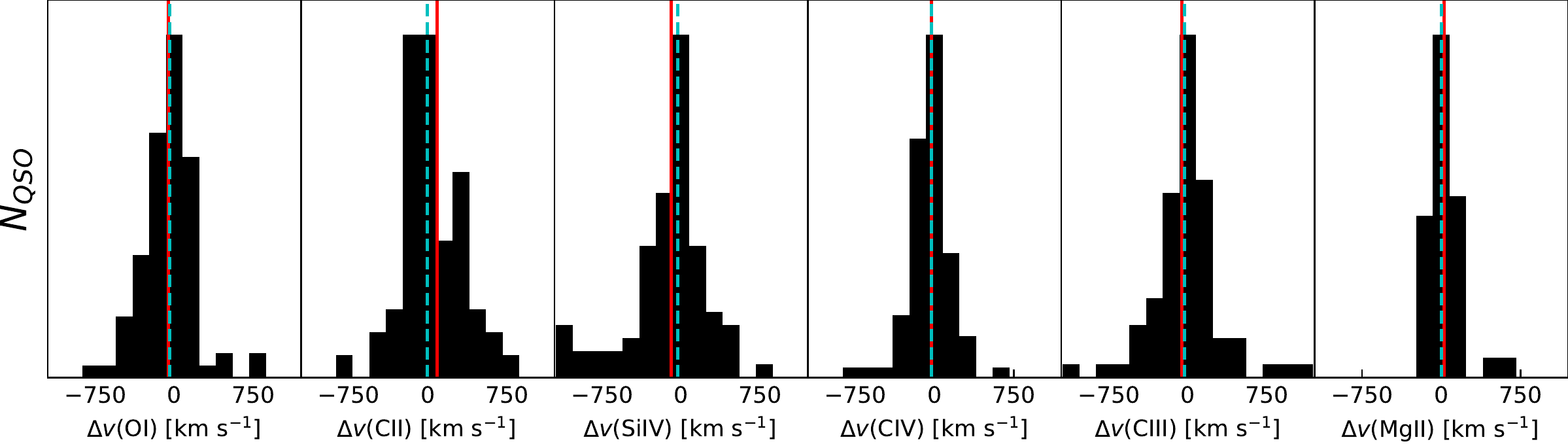}
    \caption{The impact of resolution degradation for the XQ100 spectra from $R\simeq 6000$ to $R\simeq 2000$ on derived BEL-based redshifts. The very tight distribution of velocity shifts  shows how such a spectral resolution decrease does not affect the performance of \emph{QUICFit} and does not bias the retrieved lines. The mean (red continuous line) and median (dotted blue line) offsets are much smaller than the standard deviation of the offset distribution for all species (see further Table. \ref{tab:resolution_dv_impact} for the tabulated value of the mean, median shift and the standard deviation of each distribution). }
    \label{fig:resolution_histograms}
\end{figure*}

\begin{figure}
    \centering
    \includegraphics[width=0.5\textwidth]{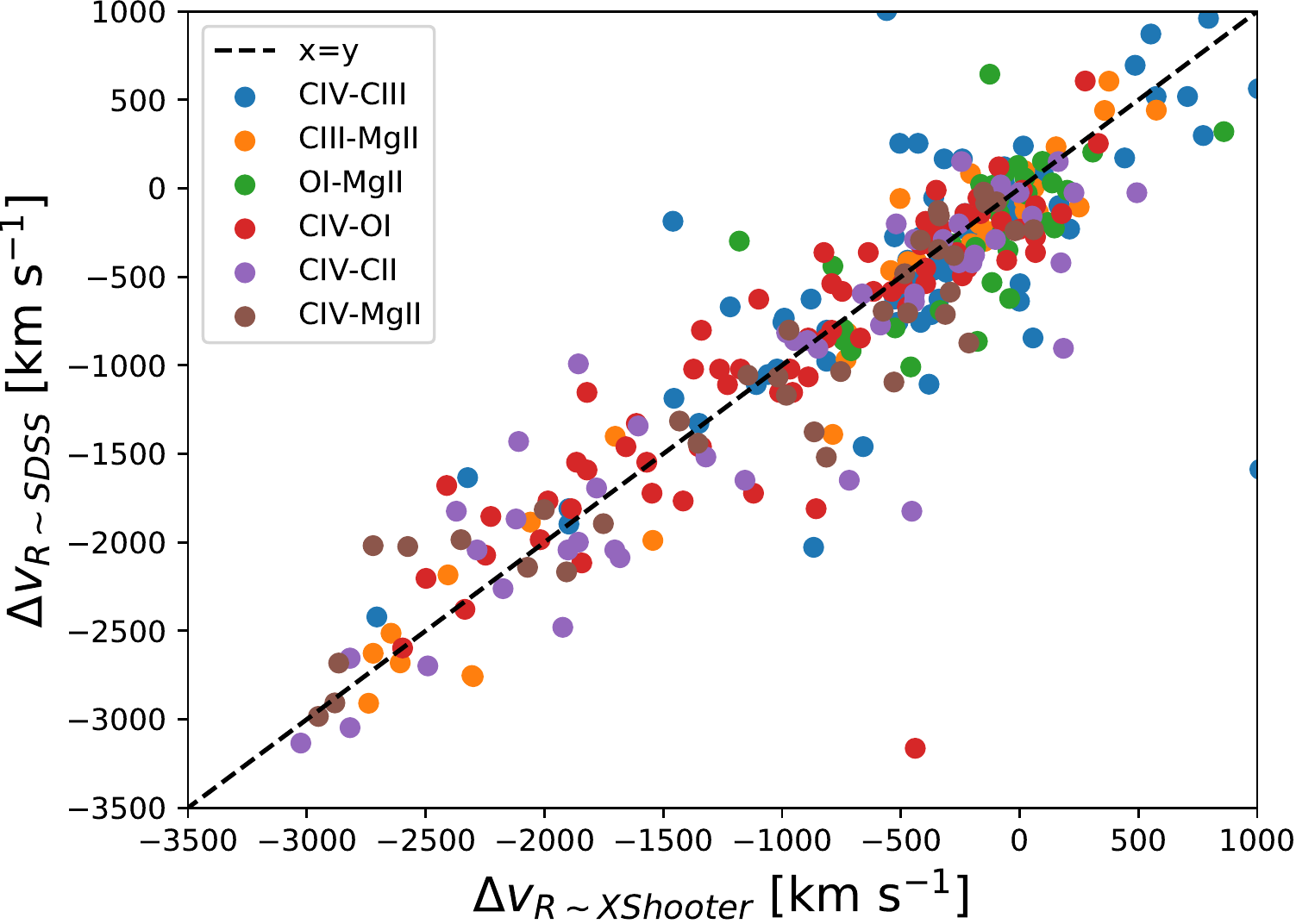}
    \caption{The impact of resolution degradation for the XQ100 spectra on derived BELs relative velocity shifts. Spectral resolution does not affect the performance of QUICFit as the figure demonstrates here the absence of bias towards higher or lower velocity shifts, despite a moderate spread (see Table. \ref{tab:resolution_dv_impact} for the tabulated values of the mean, median shift and $1\sigma$ error)}
    \label{fig:resolution_dv_XQ_rebin}
\end{figure}

In order to address this point, we have degraded the XShooter spectra to SDSS-like resolution by re-binning them by a factor 3-5 (depending on the XShooter arm and the redshift of the object) and re-fitted the degraded quasar spectra with \textit{QUICFit}. We demonstrate qualitatively the resilience of the fitting method by comparing the BEL spline fits recovered from the high- and low-resolution spectra in Figures \ref{fig:stacked_residuals} and \ref{fig:continuum_res_change}. The recovered continua and residuals for \oone, \ctwo,\sifour\, and \cthree\, are essentially identical. Nonetheless, the recovered continua for \cfour\, and \mgtwo\, is slightly lower at the peak of the BEL. As expected, the lower resolution smoothes out sharp features such as the line peaks and narrow absorbers. This causes the residuals of the low-resolution continuum to be on average $1.0\sigma-1.5 \sigma$ lower at the peak value. However the peak dampening of \cfour\, is highly symmetric, while \mgtwo\, presents a small offset to redder values. We now quantitatively qualify the potential small bias induced in the final measurement of the line-based redshifts and subsequent relative velocity shifts.
We compare the redshifts derived from the fitted BELs at low and high resolution and we find that the line-based redshifts are not biased when the resolution is lowered (Fig. \ref{fig:resolution_histograms} and \ref{fig:resolution_dv_XQ_rebin}). The average difference between high- and low- resolution is smaller than natural intrinsic scatter of the measured redshifts by a factor of $5-10$ (see Table \ref{tab:resolution_dv_impact}). Specifically, the mean redshift error due to the resolution decrease is $<100\text{ km s }^{-1}$, whereas the standard deviation is of the order of $\sim 400\text{ km s }^{-1}$.
\begin{table}
    \centering
    \begin{tabular}{lrrr|rrr} \hline
 & \multicolumn{3}{|c|}{Rebinned}   & \multicolumn{3}{|c|}{SDSS}   \\ 
Line & $\overline{\Delta v}$ & $\langle \Delta v \rangle$  & $\sigma_{\Delta v}$& $\overline{\Delta v}$ & $\langle \Delta v \rangle$  & $\sigma_{\Delta v}$ \\ \hline
\oone & $-56$ & $-44$ & $250$ & $-73$ & $-38$ & $349$ \\
\ctwo & $90$ & $-6$ & $538$ &$-25$ & $-6$ & $179$ \\
\sifour &$-96$ & $-33$ & $451$ & $-53$ & $-11$ & $407$ \\
\cfour & $-30$ & $-27$ & $537$  & $-30$ & $-16$ & $131$ \\
\cthree & $-57$ & $-30$ & $474$&$36$ & $19$ & $236$ \\
\mgtwo & $25$ & $-0$ & $166$ &$-35$ & $-19$ & $209$ \\
    \end{tabular}
    \caption{Median ($\overline{\Delta v}$), mean ($\langle \Delta v \rangle$) and standard deviation ($\sigma_{\Delta v}$) of the velocity error $\Delta v$ on of the recovered peak of each BEL with respect to the original value derived from XQ100 spectra.  The columns one to four (`Rebinned') give the values when the resolution is lowered to SDSS-like levels by rebinning the data whereas the next three (`SDSS') show the comparison between redshifts retrieved from XShooter and SDSS spectra of the same quasars. All values are provided in km s$^{-1}$.}
    \label{tab:resolution_dv_impact}
\end{table}


\begin{figure*}
    \centering
    \includegraphics[width=\textwidth]{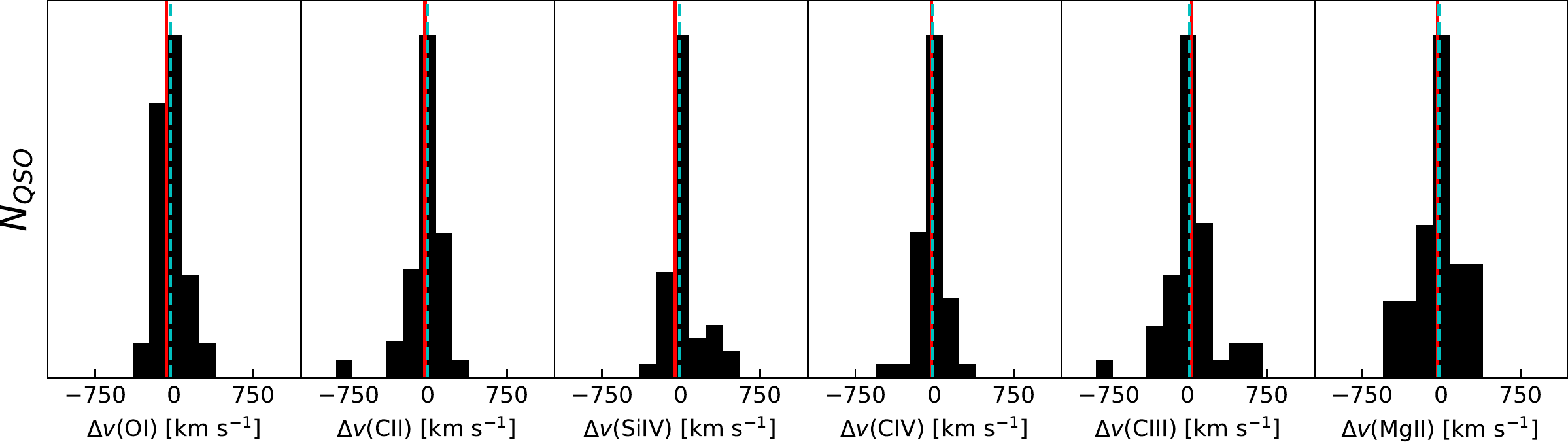}
    \caption{Comparison between derived BEL-based redshifts from XShooter and SDSS spectra of the same quasars. The very tight distribution of velocity shifts shows how a decrease in spectral resolution and different instrumental setups do not affect the performance of \emph{QUICFit} and do not bias the retrieved lines. The mean (red continuous line) and median (dotted blue line) offsets are much smaller than the standard deviation of the offset distribution for all species (see further Table. \ref{tab:resolution_dv_impact} for the tabulated values of the mean, median shift and the standard deviation of each distribution).  }
    \label{fig:resolution_histograms_XQ_SDSS}
\end{figure*}

\begin{figure}
    \centering
    \includegraphics[width=0.5\textwidth]{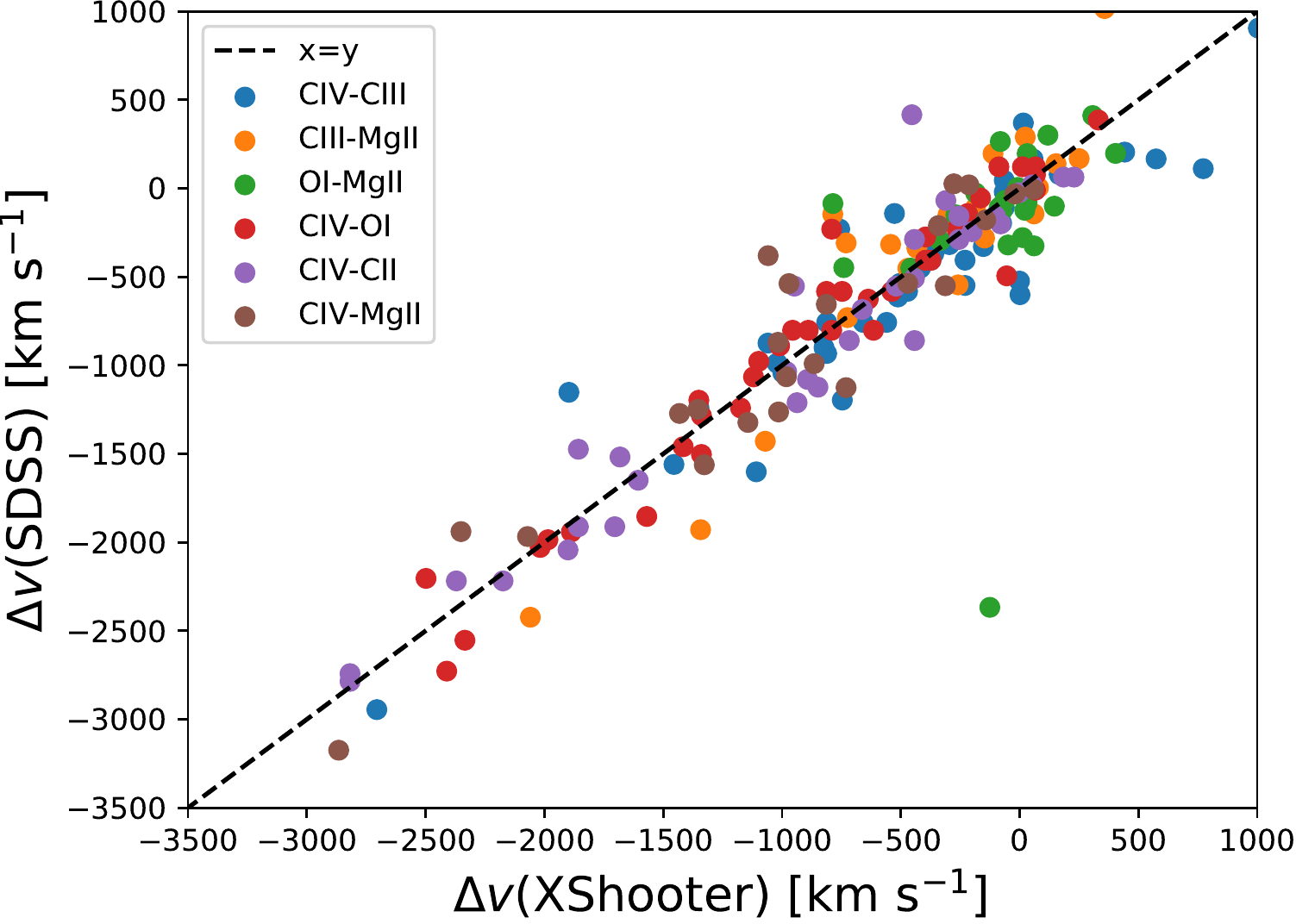}
    \caption{Comparison between derived BELs relative velocity shifts from XShooter and SDSS spectra of the same quasars. Spectral resolution does not affect the performance of QUICFit as the tight spread demonstrates here the absence of bias towards higher or lower velocity shifts (see Table. \ref{tab:resolution_dv_impact} for the tabulated value of the mean, median shift and $1\sigma$ error)}
    \label{fig:resolution_dv_impact_XQ_SDSS}
\end{figure}

To ensure that a change in both resolution and instrumental setup does not impact the measurement of the line-based redshift and relative velocity shifts, we apply our method to the $53$ XQ100 quasars observed with the SDSS spectrograph. We fit the quasar continuum with \emph{QUICFit} directly on the SDSS spectra and retrieve all possible BELs when they fall in the spectrograph coverage. We present as before the scatter in the measurement of the line-based redshifts on Figure \ref{fig:resolution_histograms_XQ_SDSS}  and the subsequent impact on the derived velocity shifts on Figure \ref{fig:resolution_dv_impact_XQ_SDSS}. The values of the mean, median and standard deviation of the line-based redshifts errors are presented in Table \ref{tab:resolution_dv_impact} alongside the previous results comparing the original XQ100 XShooter spectra to the degraded resolution version. As expected, we find no further bias by comparing XShooter and SDSS spectra than by degrading XShooter spectra. The standard deviation of the velocity shift difference between the XShooter and SDSS is often smaller than for the previous case, which we attribute to being pessimistic whilst degrading the resolution. We find no bias either in the line-based redshifts or the derived BELs relative velocity shifts. We are thus confident that our spline fitting and peak retrieval method will not bias the measured relative velocity shifts between broad lines and is independent of resolution, instrument and observational setup and date.

\subsection{Comparing \emph{QUICFit} with PCA and Gaussian template fitting methods}

\begin{figure}
    \centering
    \includegraphics[width=0.5\textwidth]{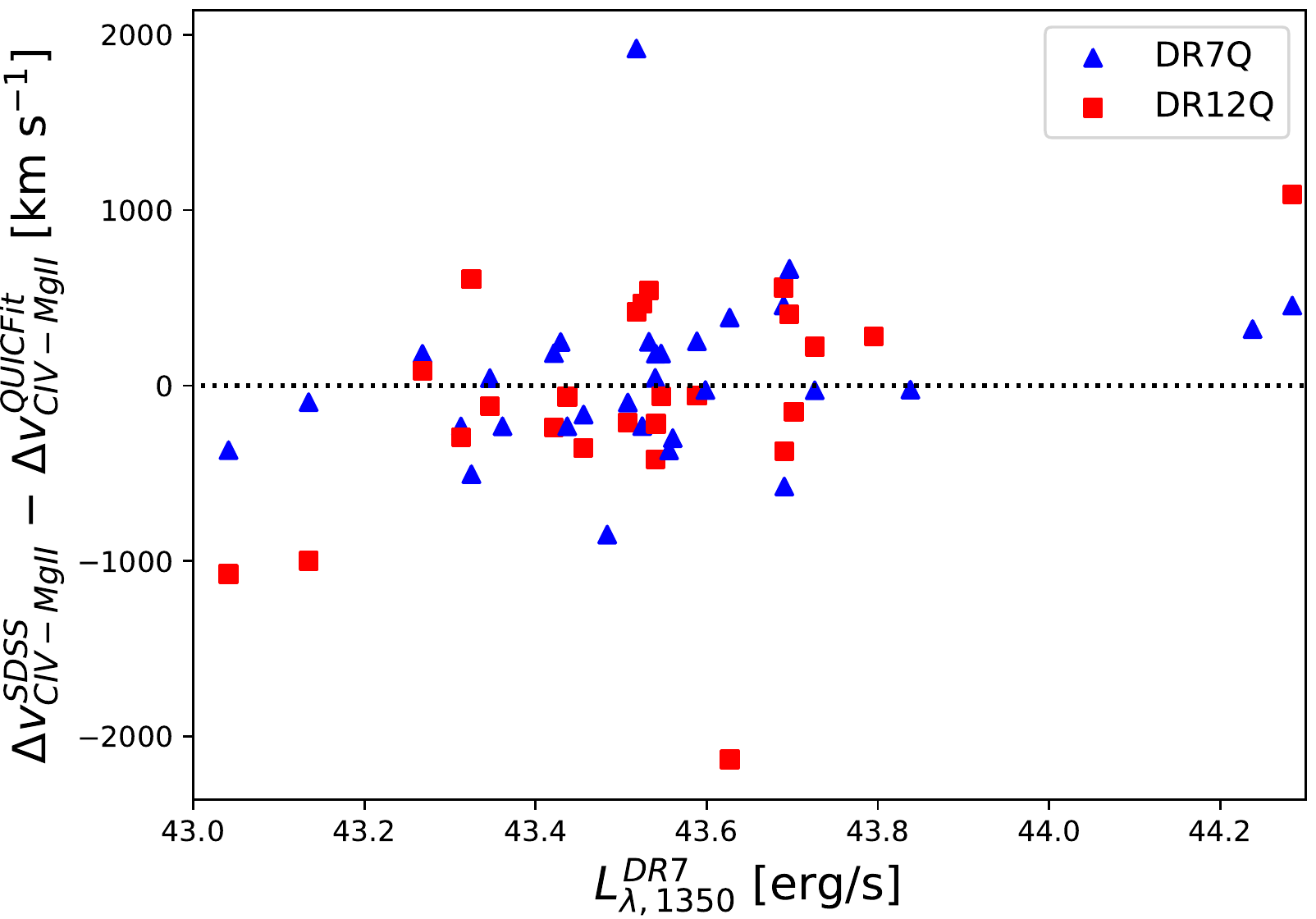}
    \caption{Comparison of the velocity shifts of \cfour\, with respect to \mgtwo\, for our SDSS subsample (Table \ref{tab:samples}). We compare the redshifts derived from Gaussian fits (DR7, blue dots) and PCA template (DR12, red square) to the QUICFit solution when these DR7Q and DR12Q provide line redshifts for both \cfour\, and \mgtwo. Although the trend is not significant, the DR12Q PCA method seems to underestimate slightly the blueshifts of more luminous quasars. In fact, the bias of the PCA towards smaller blueshifts for small \cfour\, equivalent width, which in turn anti-correlate with luminosity, was already pointed out by \citet{Coatman2016}. We find no (potential) systematic trend with redshift. }

    \label{fig:comparison_DR12_DR7}
\end{figure}

\emph{QUICFit} provides a resolution- and instrument- resilient method to fit the UV continuum and BELs of quasar continua. In this section, we compare line-based redshifts derived with our method to PCA and Gaussian fitting techniques used in previous studies. In doing so, we can also check that a change in fitting method does not bias the measurement by examining XQ100 quasars also observed in SDSS.  SDSS DR12 is too large a sample to be fitted by hand, and the DR12 quasar catalogue only contains redshifts for \cfour, \cthree\,  and \mgtwo\, \citep{Paris2017}. These redshifts are derived from the peak of the PCA template \citep{Paris2012}. DR7 redshifts, however, are derived using a Gaussian template fitting method \citep{Shen2011}. Both methods have their drawbacks and advantages, but they agree on the average \cfour-\mgtwo\, velocity shift at $1\lesssim z \lesssim 3$. In order to harness the statistical power of the DR7Q/DR12Q catalogues which measured \cfour-\mgtwo\, velocity shifts, we have to assess the possible difference between our measurements and the SDSS DR7/DR12 line-based redshifts. In order to do so, we compare the velocity shift derived from our method to the values provided by DR7Q and DR12Q where applicable in our luminosity-matched SDSS sample (Fig. \ref{fig:comparison_DR12_DR7}). We show that there is no significant bias between our values, DR7Q and DR12Q values.  The standard deviation between the different methods is however quite large. We find that the velocity shift difference between our results and DR7Q is ($\Delta v = (-83 \pm 635) \text{ km s} ^{-1}$ and $\Delta v = (47\pm  475) \text{ km s} ^{-1}$ ($\sim 500 \text{ km s}^{-1}$) for DR12Q. In both cases, the standard deviation is greater or consistent with the typical error expected from a change in resolution or a different instrumental setup as found in Section \ref{sec:methods_res}. 

\begin{figure}
    \centering
    \includegraphics[width=0.45\textwidth]{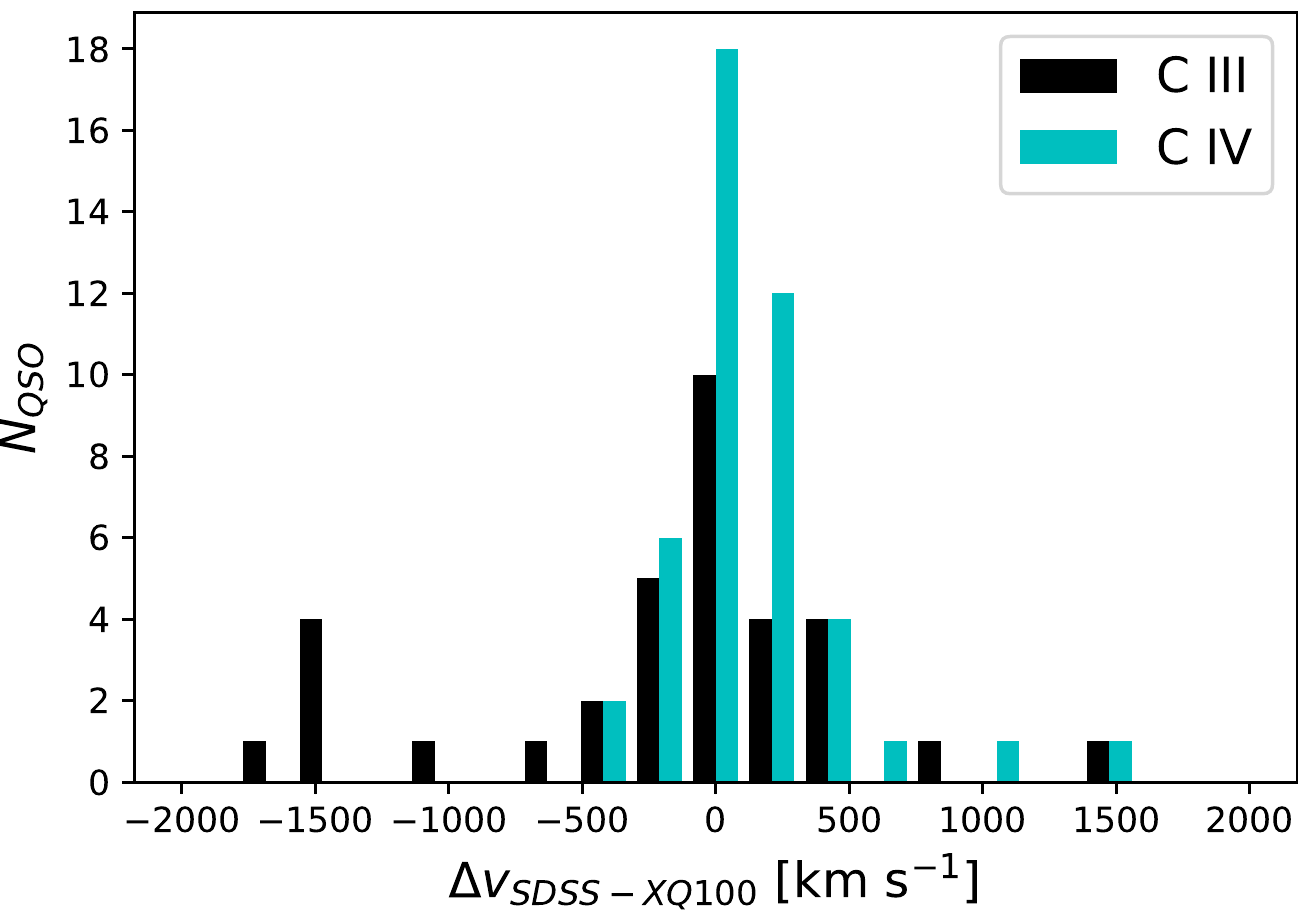}
    \caption{Comparison between the line-based (\cthree, \cfour) redshift derived from our method on the high-resolution XQ100 quasars and the tabulated values in the DR12Q catalogue for the overlapping low-resolution SDSS quasars. The methods give redshift in agreement with a typical spread of $\sim 500$ km s $^{-1}$. BELs presenting a difference of $\gtrsim 500$ km s $^{-1}$ between our solution and the DR12Q tabulated values are presented Figures \ref{fig:catastrophic_SDSS_XQ100_CIV} and \ref{fig:catastrophic_SDSS_XQ100_CIII}.}
    \label{fig:comparison_SDSS_XQ100}
\end{figure}

Finally, we can check that a concurrent change in resolution, SNR, instrument, and fitting method does not affect the measurement. For quasars present in both XQ100 and DR12Q, we compare the \cfour- and \cthree-derived redshifts from DR12Q to redshifts computed from our spline fits to the XShooter spectra. Because XQ100 contains quasars  at $3.5<z<4.5$, the corresponding SDSS observations do not cover \mgtwo. We present on Fig. \ref{fig:comparison_SDSS_XQ100} the distribution of the two methods' redshifts differences. We find that our solutions are similar to the SDSS ones with the usual $\sim 500$ km s$^{-1}$ spread for the two BELs species. The median shift, mean shift and the std deviation of the velocity shift distribution are $\overline{\Delta v_{\cfourmath}} = 53 $ km s $^{-1}$, $\langle \Delta v_{\cfourmath} \rangle = 90 $ km s $^{-1}$, $\sigma_{\Delta v_{\cfourmath}}  = 392 $ km s $^{-1}$, and  $\overline{\Delta v_{\cthreemath}} =  -7 $ km s $^{-1}$, $\langle \Delta v_{\cthreemath} \rangle = -178.6 $ km s $^{-1}$, $\sigma_{\Delta v_{\cthreemath}}  = 652 $ km s $^{-1}$ respectively. There are $\sim 10$ \% of spectra presenting shifts above $500$ km s$^{-1}$ for \cfour\, and $\sim 30$ \% for \cthree. 

For these `catastrophic' outliers, we plot their SDSS spectra, the XShooter spectra and the DR12Q and QUICFit line-based redshift solutions on Figures \ref{fig:catastrophic_SDSS_XQ100_CIV} and \ref{fig:catastrophic_SDSS_XQ100_CIII}. Upon further inspection, we find that in most cases PCA seems to fail when multiple narrow absorptions, resolved with XShooter but not SDSS, artificially damp one wing of the emission. One could argue that PCA actually recovers the original feature because it is trained on unabsorbed spectra. However, the PCA emission line redshift solutions are determined by fitting a few eigenspectra to each UV broad line individually \citep{Paris2012}. The PCA emission line redshifts are thus not derived from the overall spectral PCA fit and thus cannot reconstruct the `true' emission from other features but rather perform a local best fit to the the emission shape. Another feature that the PCA seems unable to capture is a prominent contribution of \sithree\,$\lambda 1892$ \AA\, to the \cthree\, complex. 

\section{Relative velocity shifts of broad UV emission lines} \label{sec:results}

We have demonstrated that our method is suitable for application to a large  compilation of quasars of different SNR, resolution and redshift. We have also shown that our method is comparable to two main methods for deriving line redshifts (PCA templates or Gaussian fitting), with an intrinsic spread between methods of $\Delta v \sim 500 \text{ km s} ^{-1}$. We can now harness the statistical power from our extended sample as well as our extended number of studied BELs.

\subsection{Single species broad lines: \oone, \ctwo, \cfour\, and \mgtwo}

\begin{figure*}
    \centering
    \includegraphics[width=0.7\textwidth]{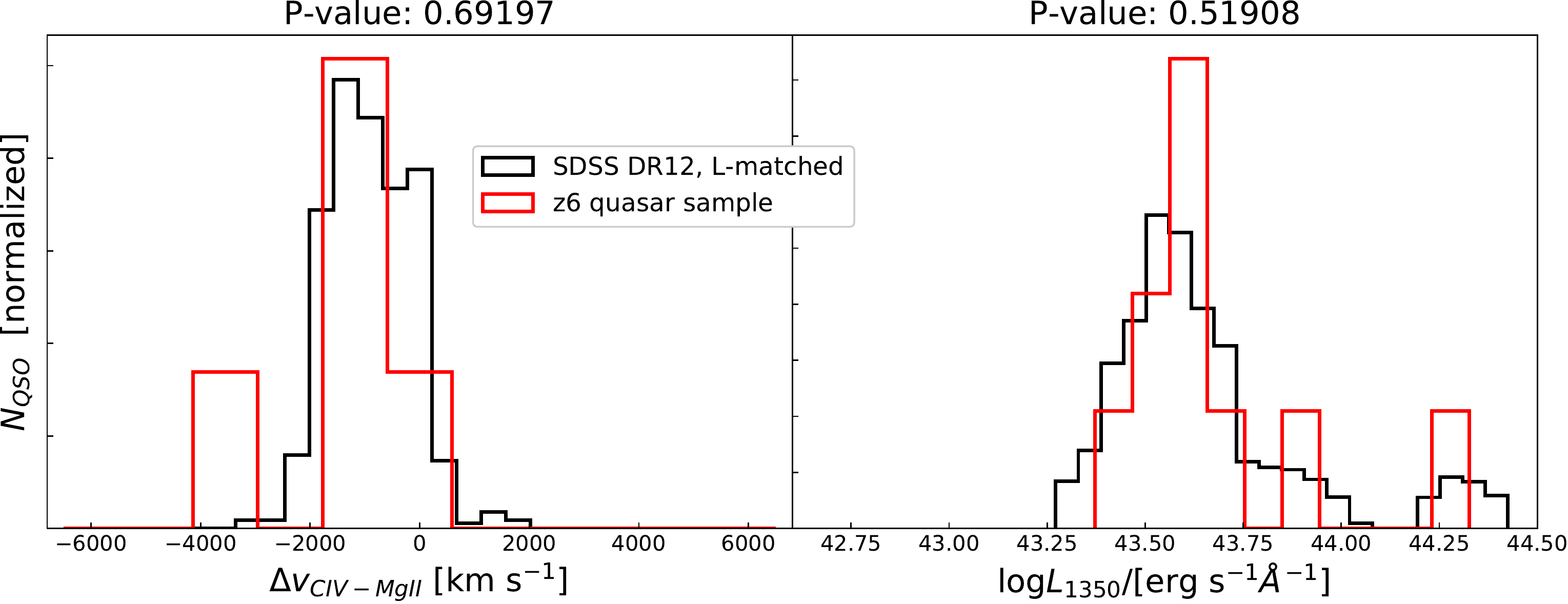} \\
    \vspace{0.2cm}
    \includegraphics[width=0.7\textwidth]{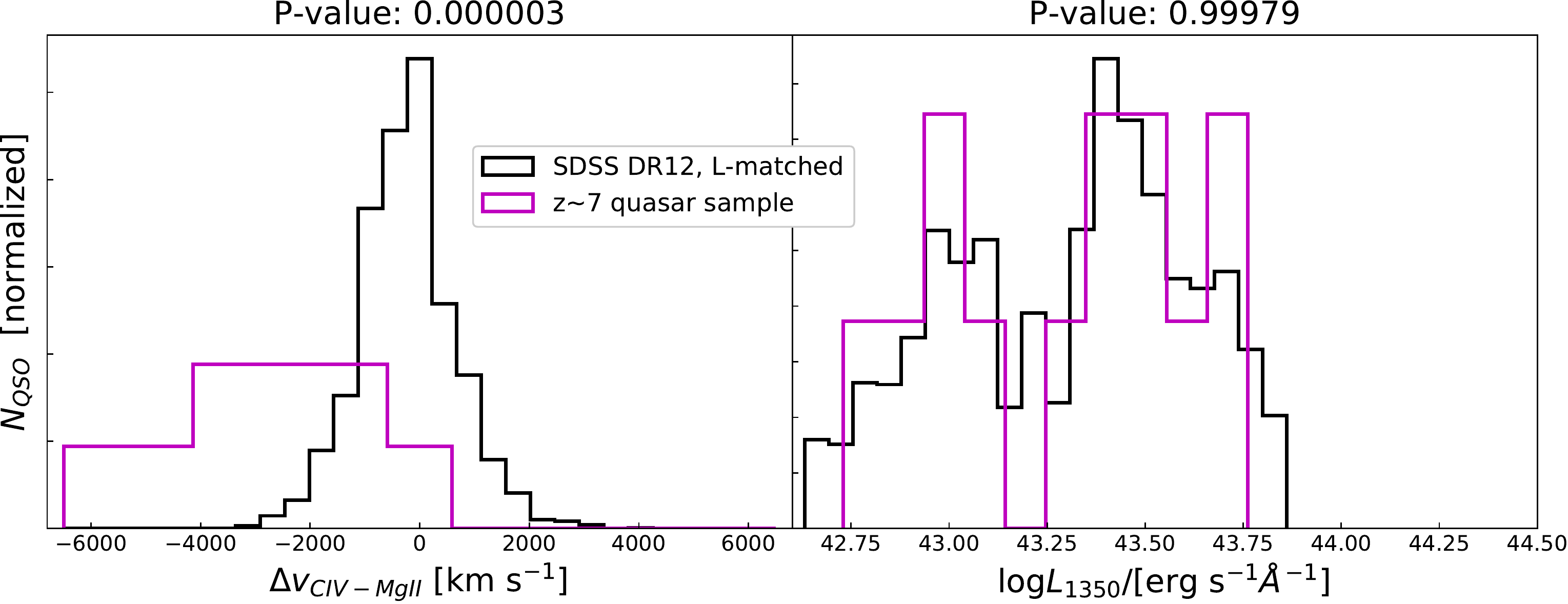}
    \caption{Comparison of the different \cfour-\mgtwo\, velocity shifts between the z6 and z7 samples and their DR12Q luminosity-matched samples. The p-value of the double-sided KS test between the test and the control samples is shown above each histogram. The z7 quasars have significantly higher \cfour-\mgtwo \, blueshifts than their luminosity-matched SDSS counterparts, whereas the z6 have comparable \cfour \, blueshifts to the lower-redshift quasars, in agreement with \citet{Shen2019}.  }
    \label{fig:histograms_CIV}
\end{figure*}

We start with the relative velocity shift of \cfour\, with respect to \mgtwo, as it is the focus of most previous studies. We note that only a fraction of our z6 and z7 quasars can be used for this analysis, as it requires good coverage of both the \mgtwo\, and \cfour\, regions.  The GMOS wavelength coverage does not allow us to detect the \mgtwo\, line, therefore we discard the GGG sample for this particular analysis. We present on Figure \ref{fig:histograms_CIV} the results for both the z6 and z7 samples compared to appropriate DR12Q control samples. We find that, when luminosity is matched, there is a significant evolution of the shifts at $z\gtrsim 6-6.5$. Specifically, the mean \cfour\, blueshift evolves from $\sim 1000 \text{ km s} ^{-1}$ at $z\lesssim 6$ to $\sim 2500 \text{ km s} ^{-1}$ at $z\sim 7$, as already reported by \citet{Mazzucchelli2017}. The double-sided KS-test, which tests whether two samples are drawn from the same different distribution, gives no evidence ($p=0.69$) for the z6 blueshifts being different from their luminosity-matched DR12Q samples. This result is similar to that reported by \citet{Shen2019} . However, the p-value for the z7 sample ($p= 3\cdot10^{-6}$) is clear evidence for evolution compared to luminosity-matched SDSS objects. The evolution of the \cfour\, line is also clearly visible in  the stacked spectra of all our samples (Fig. \ref{fig:mean_spectra}), despite the intra-object spread strongly smoothing the feature in the stacked spectrum.

\begin{figure*}
    \centering
    \includegraphics[width=0.8\textwidth]{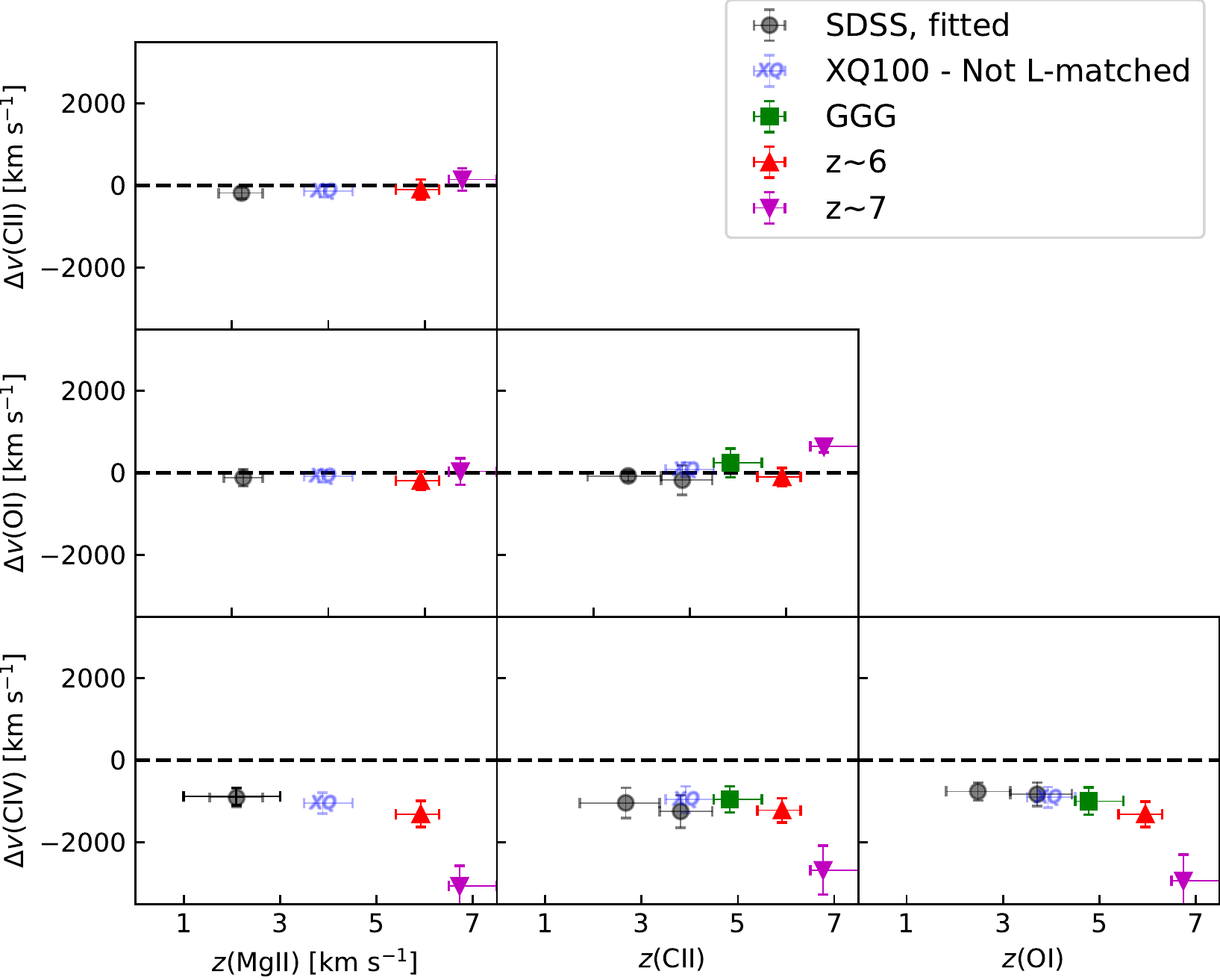}
    \caption{Velocity shifts of rest-frame UV BELs across redshift. The errors are computed by bootstrapping using samples size of the high-redshift sample. The \oone, \ctwo\, and \mgtwo\, BELs do not blueshift with respect to one another at all redshifts. However, the \cfour \, line is increasingly more blueshifted with respect to these lines at high redshift.}
    \label{fig:evol_small}
\end{figure*}

We now turn to the relative velocity line shifts for four BELs (\oone, \ctwo, \cfour, \mgtwo) in all our different samples. In order to do so, we compute the mean relative shift for every pair of species in all samples, which we show on Figure \ref{fig:evol_small}. The errors are computed by boostrapping using $1000$ samples of the size of the smallest sample (z6, $N=11$) to make them comparable across redshift and samples. The low-ionisation lines display no significant velocity offsets either between themselves at any redshift. The small relative blue- or red-shift of the lines are all consistent with bootstrapping errors. These low-ionisation line are not predicted to shift with respect to each other, therefore an absence of relative shifts is a good indicator of an unbiased measurement of the line shifts. 

The \cfour\, line, however, is markedly blueshifted from all the lower-ionisation lines (\oone, \ctwo, \mgtwo) at all redshift. The mean blueshift significantly increases at $z\gtrsim 6.5$ by a factor $2.5$, irrespective of the choice of low-ionisation species used as a reference to measure the shift of \cfour. This change is indicative of different environments for the emission (and possibly the absorption) of \cfour\, and low-ionisation lines.
It is probably linked to a dichotomy between high- and low- ionisation lines given that they are thought to originate in different parts of the BLR. To test this hypothesis, measuring the mean blueshift of another high-ionisation line is needed. However, the only other high-ionisation line in the rest-frame UV is N{\small ~V} $\lambda 1240$ \AA, which is often blended with Lyman-$\alpha$ and low-ionisation Si{\small ~II} $\lambda 1260$ \AA . Moreover Lyman-$\alpha$ is increasingly absorbed at high-redshift,
making the deblending more difficult. Another potential high-ionisation line, He {\small ~II} $\lambda 1640$ \AA\, is often too faint to be detected in most high-redshift quasars. Hence, spectroscopy of the rest-frame optical will be needed to conclude whether this blueshift is a particularity of the \cfour\, line or a dichotomy between low- and high-ionisation lines.

\subsection{Broad lines complexes: \sifour\, and \cthree}

We have left aside the \sifour\, and \cthree\, broad lines from the previous analysis. These two intermediate-ionisation lines are known to be complexes: \sifour \,$\lambda  1397$ \AA\, is blended with O{\small ~IV} $\lambda 1402$ \AA\, whereas \cthree\, $\lambda 1909$ \AA\, is blended with \althree\, $\lambda 1857$ \AA\, and \sithree\, $\lambda 1892$ \AA. 
These complexes could affect the measurement and the interpretation of any potential velocity shifts, especially since the relative line strengths of the blended emission lines appear -- qualitatively -- to be evolving with redshift.
Figure \ref{fig:mean_spectra} illustrates this effect: the O{\small ~IV} line disappears between $z\sim 5$ and $z\sim6$, whereas the  \sifour\, BEL increases in strength at high-redshift, even though the samples are luminosity-matched. 
Notwithstanding this complication, we attempted to measure the velocity shifts of these complexes by taking the red/blue peak of the complex accordingly when the feature was double-peaked, and the overall peak of the complex when the lines were blended (see Section \ref{sec:methods}). 

We find no evidence of velocity shifts of \cthree\, and \sifour\, relative to low-ionisation lines, and the \cfour\, line is seen to blueshift with respect to these lines in the same manner as to single low-ionisation lines (see Fig. \ref{fig:evol_all}). This result can be interpreted in two ways. If we were successful at disentangling the complexes, our findings indicate that \cthree\, and \sifour\, are not blueshifted with respect to low-ionisation lines at any redshift.
Even if we were not able to satisfactorily separate the blending, our results still provide some constraints on the velocity shifts of \cthree\, and \sifour. As \cthree\ is dominant over \althree\, and \sithree\ at late times and those blended lines have shorter wavelengths, a relative strengthening of \althree\, and/or \sithree\ could only lead the peak of the blend to blueshift. Redshifting of the complex peak would then be attributable to \cthree. Since neither is seen, we can plausibly conclude that on average \cthree\ does not redshift with respect to low-ionisation single species. Moreover, any intrinsic blueshift would need to be carefully masked by evolution of the other blended lines in order to go unnoticed. The situation is reversed for \sifour, which is blended with lines of longer wavelength than itself. The lack of velocity shift strongly suggests no blueshifting of \sifour, with the only alternative being a simultaneous equal shifting of O{\small ~IV} coupled with a reversal in line strength balance. 

Previous studies have already pointed at the lack of a strong Baldwin effect and blueshifts of \sifour\, \citep{Osmer1994, Richards2002}.
If the similar lack of blueshift at high-$z$ is confirmed, this would imply that \cfour\, is seen to blueshift with respect to both low- and intermediate-ionisation lines, suggesting a different origin of the line within the BLR. 


\section{Discussion}\label{sec:discussion}
\subsection{Comparison to previous works}
The most striking result of this analysis is the rapid evolution in the velocity blueshift of \cfour\, in the restricted redshift range $5<z<7$ depicted in Figure 12. Possible evolutionary trends in velocity shifts, specifically in the blueshift of \cfour\, with respect to \mgtwo, have been previously examined by several authors. \citet{Shen2019} found no significant evolution to $z\sim 6$ consistent with our results over $1.5\lesssim z\lesssim 6$. If we restrict an evolutionary test to luminosity-matched samples drawn from the z6 and SDSS datasets, the probability that these are drawn from two different distributions has an inconclusive p-value of $0.69$ (see Fig. \ref{fig:histograms_CIV}). Although the mean velocity offset at $z\simeq 6$ is slightly larger ($\Delta v_{\cfourmath-\mgtwomath} = (-1400 \pm 334) \text{ km s}^{-1}$), given the uncertainty due to the small sample size there is no convincing case for evolution.

\begin{figure}
   \centering
   \includegraphics[width=0.5\textwidth]{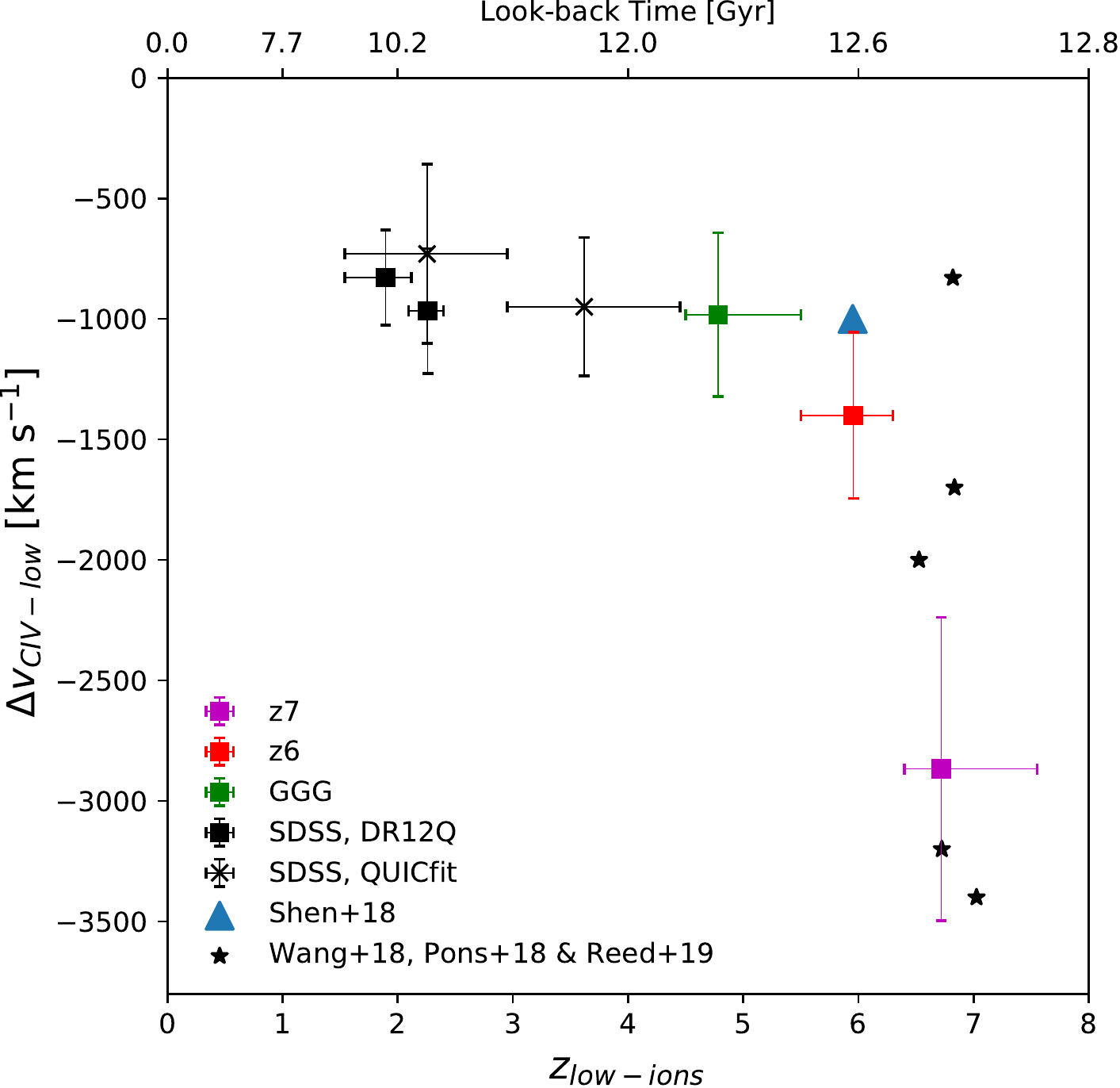}
   \caption{Velocity shifts of \cfour\, with respect to the grouped low-ionisation (\oone, \ctwo, \mgtwo) rest-frame UV quasar lines over redshift. The errors are computed by bootstrapping over samples with the size of the smallest sample (z6) to make the errors comparable. The  apparent drift in the blueshifts of the SDSS fitted subsample is due to the reference low-ionisation lines which include only \mgtwo\, at low-redshift, and then gradually move over to \oone and \ctwo\, at high-redshift. The $\sim 500$ km s $^{-1}$ velocity shift is consistent with the small relative shifts of these low-ionisation lines (see Fig. \ref{fig:evol_small}). We also add the reported values of $\Delta v_{\cfourmath-\mgtwomath}$ of the \citet[][blue triangles]{Shen2019} sample and recent individual high-redshift quasars of similar luminosities \citep[black stars, ][]{Wang2018,Pons2019,Reed2019}. }
   \label{fig:evol_binned}
\end{figure}

However, with the addition of the $z\sim 7$ quasars from \citet{Mazzucchelli2017} and \citet{Banados2018}, we recover the larger mean blueshift of \cfour\, found by these authors. Since we have applied the same method to different samples over the redshift range $1.5 \leq z \leq 7.5$, we argue this evolution cannot be caused by resolution or other instrumental effects. Our method also allows us to extend the analysis by substituting \oone\, or \ctwo\, for \mgtwo\,. Given the similarity of the trends so revealed, we can aggregate over the low-ionisation lines to construct a \cfour-low-ionisation velocity shift versus redshift trend that uses all datasets (Fig. \ref{fig:evol_binned}). When aggregating the low-ionisation species, we first construct a common `systemic' redshift from the mean of all available low-ionisation line-based redshifts. We then derive the \cfour\, blueshift and compute errors by bootstrapping as before. The sharp increase in the \cfour\, blueshift between $z\sim 5-7$ is evident. We also show that the recently published \cfour-\mgtwo\, velocity shifts for individual $z\gtrsim 6.5$ quasars are in agreement with the increasing mean \cfour\, blueshift trend \citep{Wang2018,Pons2019,Reed2019}.

There has been some debate on the possibility that biases in the line-based redshift determination of SDSS quasars might remove the most blueshifted \cfour\, lines. DR7Q derives \cfour\, blueshifts up to a maximal value of $5000 \text{ km s}^{-1}$, making the comparison with some $z\sim 7$ quasars impossible. \citet{Coatman2016} showed that using a definition of the line center based on the peak emission or line centroid was strongly affecting the SDSS distribution of \cfour\, blueshifts, potentially alleviating some concerns raised with the first reports of large blueshifts in early quasars \citep{DeRosa2014}. The recent results of \citet{Reed2019} showcases the use of a currently unpublished Independent Component Analysis (ICA) determination of the SDSS \cfour\, blueshifts that identifies multiple extreme blueshifts $>4000 \text{ km s}^{-1}$ in SDSS. Despite both studies extending the range of existing SDSS \cfour\, blueshifts, the mean shift and the abundance of redshifted \cfour\, lines still does not match that for $z\sim 7$ quasars. We have tested this by independently measuring the line-base redshifts of \cfour\, and different low-ionisation BELs in about a hundred luminosity-matched SDSS quasars. This result suggests that improvements in quasar template fitting will most likely not result in a dramatic increase of the average \cfour\, blueshift at low redshift. Nonetheless, these advances will enable the selection and study of low redshift quasars with extreme \cfour\, blueshifts, which we argue below will be a crucial point to further our understanding of reionisation quasars.

\subsection{Interpreting the high-redshift increased mean \cfour\, blueshift}

The most intriguing aspect of the trend in Fig. \ref{fig:evol_binned} is that the inferred evolution occurs in less than 1 Gyr. At low redshift, discussion of the nature of \cfour\ blueshifts has centered around radiation-driven winds originating in the central regions, or the BLR, of SMBH accretion discs. The broadening of quasar UV emission lines was first modelled through primarily planar outflows, in which material is stripped from the accretion disc and accelerated outwards by X-ray radiation and/or resonant line radiation pressure from the inner regions \citep[e.g.]{Krolik1986, Murray1995}. Based on the singular behaviour of the \cfour\ BEL, \citet{Wills1993} first suggested a possible separate origin of the line in a hotter, polar wind component (see also \citealt{Denney2012}). Later models showed how gas could be entrained at a steeper angle from more central regions (e.g. via increased magnetisation of the outflows, \citealt{Proga2000, Elvis2000}) and even give rise to outflows which are significantly polar by entraining gas from a dusty torus, rather than a thin accretion disc (e.g. \citealt{Gallagher2015}). Early results from sub-millimeter imaging have supported the possibility of significantly perpendicular winds, with detections of elongated dusty emission around AGN in Seyfert galaxies being preferentially \textit{parallel} to their ionisation cones, rather than perpendicular (\citealt{Lopez-Gonzaga2016a} and references therein). If low-ionisation and high-ionisation broad lines do indeed arise through such related but different processes along different axis, with higher-ionisation outflows being more polar, it is possible to conceive scenarios in which one evolves but not the other.

The most commonly invoked explanation for the fast evolution of \cfour\ blueshifts at early times involves (unusually) strong BLR outflows or winds to explain (extremely) blueshifted \cfour\, emission \citep[e.g.][]{Richards2011,DeRosa2014,Mazzucchelli2017}. This claim is supported by the widespread presence at $z\sim 7$ of extreme blueshifts with $\Delta v \gtrsim 3000$ km s $^{-1}$ which are exceedingly rare at $z<5$. 
However, it is unclear what mechanism could power such an increase in wind speed in early quasars, or whether this can account for the whole effect. Such extreme blueshifts are found to be quite common in some lower-redshift populations such as the WISE/SDSS selected hyper-luminous  quasars \citep[WISSH, ][]{Bischetti2017,Vietri2018} at $z\sim 2-4$,  and hence are not exclusive to $z\gtrsim 7$ quasars. 
Specifically, WISSH quasars with weaker O{~\small III} $\lambda 5007$ \AA \, exhibit larger blueshifts and lower X-ray to optical luminosity ratio, which are also found in some high-redshift quasars \citep{Banados2018}, and this trend seems to persist even in moderately bright SDSS quasars \citep{Coatman2019}. This suggests that mechanisms which increase \cfour\ blueshift are still in place at much later times, and even dominant in some populations.
Moreover, anomalously large \cfour\ blueshifts at early times do not seem to correlate with higher accretion rates -- as one might perhaps expect if they were caused primarily by anomalously strong winds.
\citet{Mazzucchelli2017} reported that their $z\gtrsim 6.5$ sample had high accretion rates near the Eddington limit and their SDSS luminosity-matched sample did as well. Finally, a wind-only model would imply that some cases of no blueshift or even redshifted lines should always occur due to orientation with respect to the observer, unless the wind is increasingly spherical. In contrast, there is a total absence of redshifted \cfour\, emission lines in luminous quasars beyond $z\gtrsim 6$. 
If increased wind speeds do power the increase in \cfour\ blueshifts, they are then most likely accompanied by a change in morphology.
This conclusion either suggests a fundamental change in the properties of quasars as they emerge from the reionisation era or perhaps some selection effect which would cause quasars with non-blueshifted \cfour\, lines to be absent at $z\gtrsim 6$ despite our attempts to create luminosity-matched samples.

\subsubsection{Orientation selection bias}
In a simple `opaque torus' model of quasars \citep[][and references therein]{Denney2012}, \cfour\, radiation originates either in the broad line region or a surrounding intermediate line region and launches fast winds perpendicular to the plane of the accretion disk (or at least significantly more polar than low-ionisation winds).  If the accretion disk is viewed side-on where the line of sight is opaque, this will result in a broad \cfour\, emission with no blueshift and a dimmed UV quasar luminosity. Viewed increasingly face-on, the \cfour\, line will be blueshifted with a shape dependent on the wind velocity profile. This geometric model explains many observational features of broad quasar emission lines, such as the consistent, but moderate average blueshifts at low redshift.

Assuming that high-redshift quasars are obscured to the same extent as those at lower redshifts, the enhanced \cfour\, velocity offsets would then likely be related to the relative youth of quasars at the end of the reionisation era. Although the $z>6.5$ quasars in our sample have already massive ($\sim10^9 M_\odot$) black holes accreting at nearly the Eddington rate \citep{DeRosa2014,Mazzucchelli2017,Banados2018}, the UV luminosity of quasars would be expected to be attenuated when they are viewed side-on, which may lead to them not being detectable at our luminosity threshold at $z\gtrsim 6.5$. In this picture, quasars with masses and accretion rates large enough to be UV-bright despite being side-on might not have yet assembled as early in cosmic time. Sampling the top of the UV luminosity distribution at any cosmic time would then unwittingly bias observations towards face-on objects with large \cfour\, blueshifts. 

\begin{figure*}
    \centering
    \includegraphics[width=\textwidth]{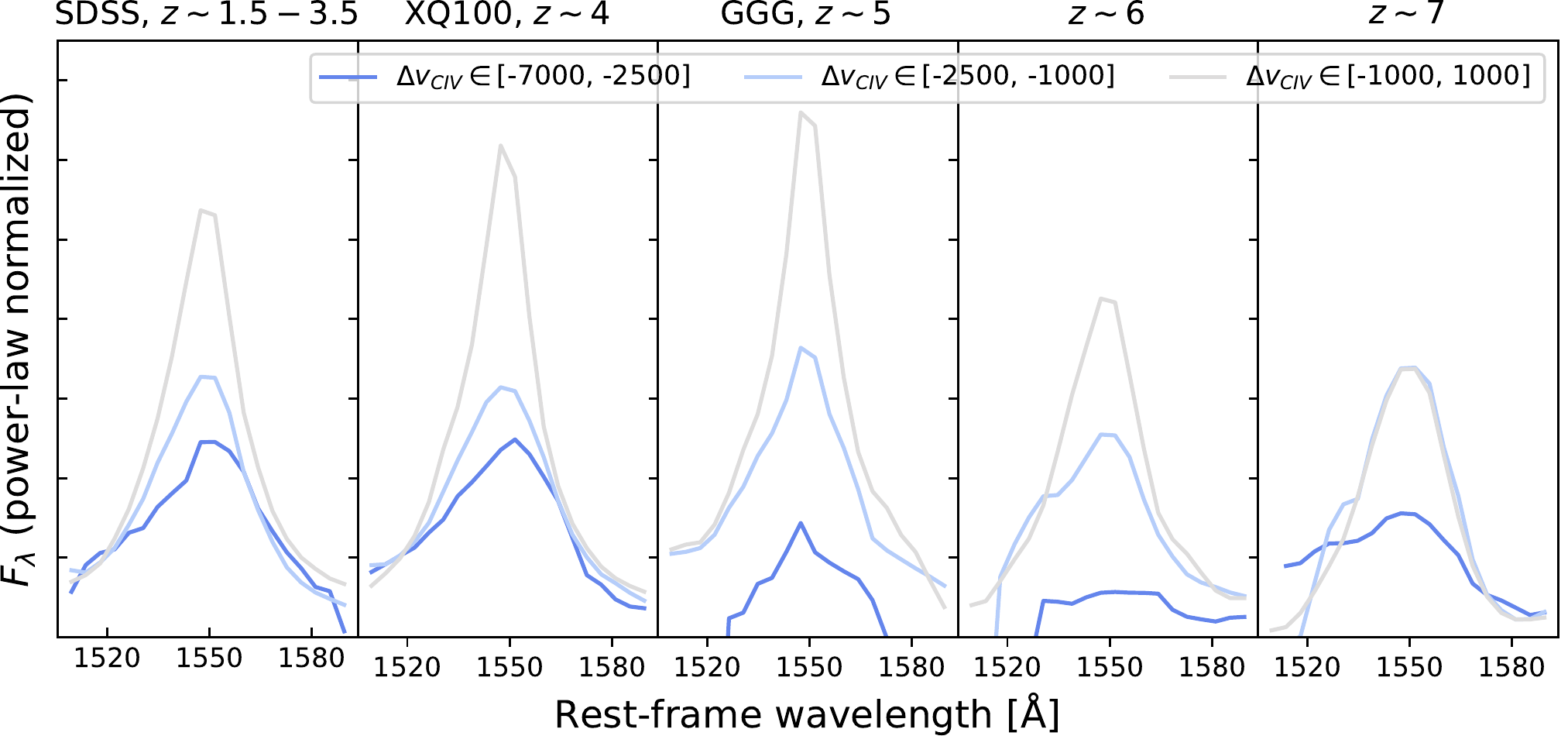}
    \caption{Profiles of the \cfour\, BEL for quasars at $1.5\lesssim z \lesssim 7.5$, stacked according by the \cfour\, blueshift (gray: low blueshifts ($<1000$ km s $^{-1}$), light blue: moderate ($1000 \text{ km s} ^{-1} \leq\Delta v_{\cfourmath}\leq 2500 \text{ km s} ^{-1}$), dark blue: extreme blueshifts ($>2500$ km s $^{-1}$) from low-ionisation lines (\oone, \ctwo, \mgtwo). When multiple low-ionisation lines are detected, we take the average redshift to compute the relative blueshift of \cfour. The evolution of the blueshifts of \cfour\, is matched by the evolution of the profiles. The emission is weaker at high-redshift, and the change is more prominent for the non-blueshifted objects. We attribute this to increased obscuration at high-redshift as the non-blueshifted \cfour\, are originating in quasars seen edge-on in our interpretation (see Section \ref{sec:discussion}).}
    \label{fig:stack_civ_dv}
\end{figure*}

If this selection bias were the only effect at work, quasar continuum reconstruction models which successfully capture the variety of low-redshift quasars should be expected to work similarly well at high-redshift. Indeed, the physical origin of the \cfour\, blueshift -- outflows and orientation -- would be intrinsically the same across cosmic time.
One might expect a range of related observational consequences, for example:\begin{itemize}
\item The most extremely blueshifted objects of a given luminosity -- being face-on -- should present roughly the same $\Delta v_{\cfourmath}$ at all redshifts. In our study, the maximum blueshift in the $z\sim 7$ sample is $2000 \text{ km s}^{-1}$ greater than our DR12Q luminosity-matched sample extremum (Figure 9). As our samples are relatively small, it is possible that larger datasets of high-redshift quasars and new fitting techniques of low-redshift quasars could still change the values of the most extreme blueshifts. 
\item The shape and strength of the highly blueshifted \cfour\, emission lines should be the same across redshift, since the evolution of the mean shift would be due only to the disappearance of quasars with non-blueshifted \cfour. We present in Fig. \ref{fig:stack_civ_dv} the mean \cfour\, line profiles of our quasars at different redshifts, stacked according to the blueshift of the \cfour\, emission compared to low-ionisation lines. It can be seen that the higher-redshift quasars show a tentatively attenuated profile and smaller equivalent widths when \cfour\, is not blueshifted, as reported in previous studies \citep[][]{Mazzucchelli2017,Shen2019}. 
\item The effect should vanish for fainter high-redshift quasars. Perhaps the best test of this `selection bias' would be to examine the \cfour\, velocity shift at early times in less luminous objects. Such a strategy is obviously very challenging. 
However, the discovery of a faint $z\sim 7$ quasar with a significant \cfour\  blueshift ($\sim 2500 \text{ km s}^{-1}$) in the Subaru High-$z$ Exploration of Low-Luminosity Quasars survey (SHELLQS) provides marginal evidence against this interpretation \citep{Matsuoka2019} while a recent small sample of six relatively faint quasars at $z\sim6.3$ show mixed results including potentially the first \cfour\ redshift with respect to \mgtwo\ at $z>6$ \citep{Onoue2019}. Clearly a larger sample of fainter, possibly lensed quasars above $z\gtrsim 6$ can test this hypothesis. 
\end{itemize}

\subsubsection{Increased obscuration}
A competing or complementary alternative is that dust obscuration is more intense in early quasars. In this case, those quasars viewed side-on will again be (even more) strongly de-selected at high-redshift, increasing the average blueshift of the \cfour\, line. \citet{Treister2011} reached the conclusion that black hole accretion is mostly obscured in the Early Universe by looking at the X-ray emission at $z\sim 6-8$ attributed to early quasars. \citet{Trebitsch2019} showed that increased quasar obscuration at high-redshift could perhaps be due to chaotic feeding and instabilities in the less mature host galaxies. The same authors also showed that early quasars accreting at higher Eddington ratios are preferentially surrounded by dense, obscuring gas and are only being visible with $M_{1450}>-23$ about $4$\% of the time. Given that the highest redshift quasars must have accreted at nearly Eddington rate for most of their life \citep[e.g.][]{Banados2018}, it is thus likely that they are more obscured than the lower-redshift ones. By selecting the most UV-bright objects at high-redshift we would then only find quasars which have both very high accretion rates {\textit{and}} are preferentially orientated, with potentially very narrow visibility channels. We speculate that chaotic feeding of the central black hole would affect the BLR and thus the BELs. Models and templates extracted from low redshift quasars would thus be expected to fare fairly poorly on high-redshift objects.

This second scenario might also explain some other intriguing recent finds in reionisation era studies. The small near-zones of some high-redshift quasars \citep{Eilers2017,Eilers2018a} could be due to intermittent obscuration of the quasar, blocking LyC photons up to $z\sim 6$, where the near-zone would start to grow steadily and \cfour\, blueshift normalize.  Depending on the type of obscuration at work, highly obscured and thus UV-faint quasars might in fact already be known as faint AGNs hosted in high-redshift galaxies showing N{~\small V} emission lines \citep[e.g.][]{Stark2017}.

Finally, both interpretations have potential consequences for the contribution of quasars to hydrogen reionisation \citep[e.g.][]{Madau2015}. In a `geometrical selection effect' scenario, only quasars which are `sufficiently face-on' can be currently detected at high redshift. Indeed, $\sim50\%$ of z7 quasars have \cfour\, blueshifts $\geq 2500$ km/s, while those make up $<15\%$ (at $2\sigma$) of the DR12Q sample. Since quasar ionising radiation output is calculated over all angles, this would imply that an average high-redshift quasar in our luminosity-matched sample contributes {\textit{less}} to the IGM photon budget than its low-redshift counterpart. We deduce this because edge-on, non-\cfour-blueshifted quasars on average are expected to have larger masses and larger accretion rates in an orientation-only model -- and those are the ones missing at early times.
By contrast, in a scenario where early quasars are more obscured than their low-redshift counterparts, it is likely that we are currently  observationally missing a large fraction.
 We would expect to find an increased number of obscured early AGN -- which are perhaps contributing to the bright end of the galaxy UV luminosity function \citep{Ono2018} or to the AGN X-ray luminosity function, which should be less obscured \citep{Giallongo2015}.
Using the rough value of $\sim4\%$ UV-bright visibility fraction from \citet{Trebitsch2019},  and if obscuration is mostly geometric, we could be underestimating the number density of early massive quasars by factors of $3-4$. This deficit could be somewhat balanced by less ionising emission escaping through smaller channels. In any case, the \cfour\, blueshift conundrum has far-reaching consequences on the role of quasars in reionisation. 

\section*{Acknowledgements}
We authors thank the anonymous referee for the thoughtful comments on the manuscript. RAM, SEIB and RSE acknowledge funding from the European Research Council (ERC) under the European Union's Horizon 2020 research and innovation programme (grant agreement No 669253). We would like to thank E. Ba\~nados and C. Mazzucchelli for useful discussions and kindly sharing their data with us. We thank M. Trebitsch for useful discussions. 

Based on observations collected at the European Organisation for Astronomical Research in the Southern Hemisphere under ESO programme(s) 084A-0550, 084A-0574, 086A-0574, 087A-0890 and 088A-0897. \\ 
Based on observations made with ESO Telescopes at the La Silla or Paranal Observatories under programme ID(s) 189.A-0424(A), 189.A-0424(B) \\

Funding for the SDSS and SDSS-II has been provided by the Alfred P. Sloan Foundation, the Participating Institutions, the National Science Foundation, the U.S. Department of Energy, the National Aeronautics and Space Administration, the Japanese Monbukagakusho, the Max Planck Society, and the Higher Education Funding Council for England. 

Funding for SDSS-III has been provided by the Alfred P. Sloan Foundation, the Participating Institutions, the National Science Foundation, and the U.S. Department of Energy Office of Science. The SDSS-III web site is http://www.sdss3.org/.

SDSS-III is managed by the Astrophysical Research Consortium for the Participating Institutions of the SDSS-III Collaboration including the University of Arizona, the Brazilian Participation Group, Brookhaven National Laboratory, Carnegie Mellon University, University of Florida, the French Participation Group, the German Participation Group, Harvard University, the Instituto de Astrofisica de Canarias, the Michigan State/Notre Dame/JINA Participation Group, Johns Hopkins University, Lawrence Berkeley National Laboratory, Max Planck Institute for Astrophysics, Max Planck Institute for Extraterrestrial Physics, New Mexico State University, New York University, Ohio State University, Pennsylvania State University, University of Portsmouth, Princeton University, the Spanish Participation Group, University of Tokyo, University of Utah, Vanderbilt University, University of Virginia, University of Washington, and Yale University.  




\bibliographystyle{mnras}
\bibliography{QSOBEL} 



\appendix
\section{Line-derived redshifts for all quasars}
\label{appendix:fits}
We present the continuum splines fitted around the broad emission lines for each quasar in this study. Panels from left to right show broad emission lines arranged by increasing rest-frame wavelength, i.e. : \oone, \ctwo, \sifour, \cfour, \cthree, \mgtwo. For each quasar and line, we show the observed flux (black), the fitted continuum splines (blue line) and the location of the peak if it satisfies the criteria of Section \ref{sec:methods} (red vertical liens). For XQ100 quasars, we also show the peak recovered from the low-resolution version of the spectra (green vertical lines) and the one derived from the SDSS spectra if the object is present in SDSS (blue dotted vertical lines). Each BEL and fitted continuum are rescaled to facilitate the inspection. We thus caution that the lines are all at a different scale except for \oone\, and \ctwo\, which are rescaled by the same amount as they are contiguous in rest-frame wavelength. For the XQ100 quasars, we also show the continuum splines fitted to the resolution-degraded flux, and the derived peak positions thereof. The quasar name and literature redshift are indicated on the left of all panels. We give only the fits for the first $6$ objects of the XQ100 sample to limit the length of this manuscript. The remaining plots are available as online supplementary material.

Eventually, we present a summary table of our measured quasar redshift based on each BEL in Table \ref{tab:all_redshifts}. The entire tables for each samples are available as supplementary material.


\begin{figure*}
    \centering
    \includegraphics[width=0.95\textwidth]{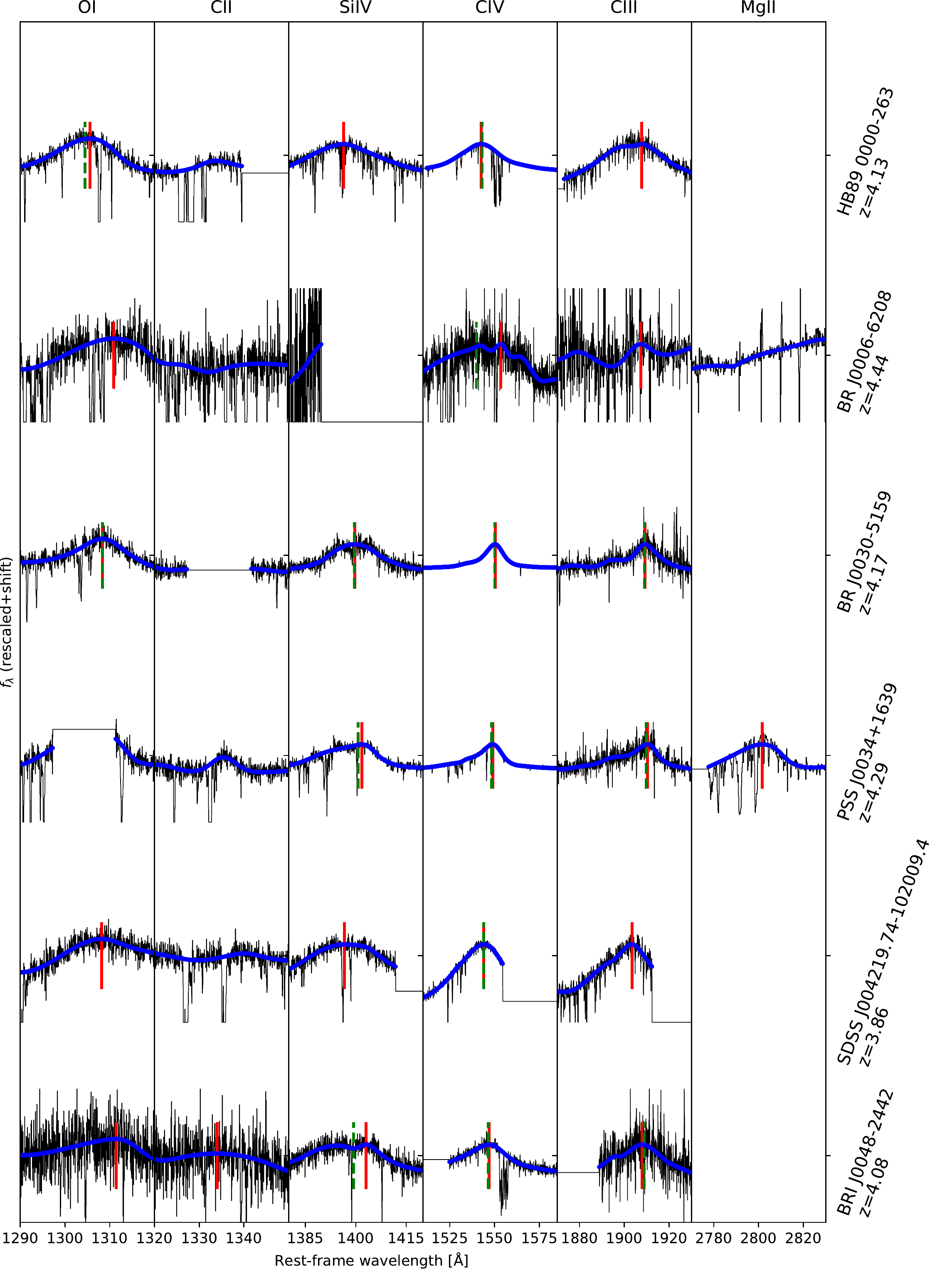}
    \caption{XQ100 BELs fits given for the six lines of interest. The observed flux (black) is shown alongside the spline fits (blue) in the rest-frame computed from the literature redshift indicated on the vertical right axis. We indicate the location of peaks used to derived BEL-based redshifts by a vertical red line. We also indicate the peak solution derived from the low-resolution version of the spectra (vertical green line) and the SDSS spectra if available (vertical dotted blue). }
    \label{fig:fits_XQ100}
\end{figure*}



\begin{table}
    \centering
    \begin{tabular}{lrrrrrr}\hline
Quasar &  $z_{\oonemath}$ & $z_{\ctwomath}$  & $z_{\sifourmath}$ & $z_{\cfourmath}$ & $z_{\cthreemath}$ &$z_{\mgtwomath}$ \\ \hline
\hline
\textbf{SDSS Quasars} & & & & & & \\ 
\hline 
5379-55986-0904 & 3.87 & 3.87& 3.87 & 3.81 & 3.84 &  \\ 
4481-55630-0178 & 2.52 &  & 2.51 & 2.50 & 2.49 & 2.52\\ 
4193-55476-0376 & 2.28 &  & 2.26 & 2.26 & 2.25 & 2.27\\ 
$\vdots$ & & $\vdots$& &$\vdots$ & & $\vdots$\\
\hline
\textbf{XQ100 Quasars} & & & & & & \\ 
\hline
HB89 0000-263 & 4.13 &  & 4.11 & 4.10 & 4.12 &  \\ 
BR J0006-6208 & 4.46 &  &   & 4.45 & 4.44 &  \\ 
BR J0030-5159 & 4.18 &  & 4.17 & 4.18 & 4.17 &  \\ 
$\vdots$ & & $\vdots$& &$\vdots$ & & $\vdots$\\
\hline
\textbf{GGG Quasars} & & & & & & \\
\hline 
SDSSJ0011+1446 & 4.96 & 4.97& 4.95 & 4.94 &   &  \\ 
SDSSJ0822+1604 &   &  & 4.47 & 4.50 &   &  \\ 
SDSSJ1043+6506 &   &  & 4.48 & 4.46 &   &  \\ 
$\vdots$ & & $\vdots$& &$\vdots$ & & $\vdots$\\
\hline
\textbf{z6 Quasars} & & & & & &\\
\hline
J0148+0600 & 5.97 & 5.96& 5.93 & 5.91 & 5.93 & 5.98\\ 
J0836+0054 & 5.80 & 5.80& 5.81 & 5.78 & 5.76 & 5.77\\ 
J0927+2001 & 5.74 & 5.76& 5.73 & 5.72 & 5.77 & 5.76\\ 
J1030+0524 & 6.30 & 6.27& 6.30 & 6.28 & 6.31 & 6.30\\ 
J1306+0356 & 6.03 & 6.02& 6.02 & 6.00 & 6.01 & 6.02\\ 
J1319+0950 & 6.10 & 6.11& 6.17 & 6.03 & 6.14 & 6.12\\ 
J0100+2802 & 6.31 &  & 6.25 & 6.24 & 6.30 &  \\ 
J0818+1722 &   &  &   &   & 5.95 &  \\ 
J1509-1749 & 6.08 & 6.11& 6.11 & 6.10 & 6.10 & 6.12\\ 
J1044-0125 & 5.78 & 5.80& 5.77 & 5.76 & 5.79 & 5.78\\ 
J0231-0728 & 5.42 & 5.42& 5.43 & 5.42 & 5.42 & 5.42\\ 
\hline 
\textbf{z7 Quasars} & & & & & & \\
\hline
J2348-3054 & 6.91 & 6.89& 6.87 & 6.85 & 6.89 & 6.89\\ 
P231-20 & 6.57 &  & 6.68 & 6.47 & 6.54 & 6.60\\ 
P167-13 & 6.50 & 6.49& 6.41 & 6.41 & 6.48 & 6.51\\ 
P036+03 &   &  & 6.45 & 6.42 & 6.50 & 6.53\\ 
J0305-3150 & 6.56 &  & 6.57 & 6.57 &   & 6.61\\ 
P183+05 &   & 6.48& 6.40 & 6.35 &   & 6.42\\ 
J1120+0641 &   & 7.09& 7.04 & 7.01 & 7.06 & 7.09\\ 
J1342+0928 & 7.56 & 7.55& 7.41 & 7.35 &   & 7.52\\ 
P247+24 & 6.45 & 6.46& 6.44 & 6.42 &   & 6.48\\ 
P338+29 &   & 6.64& 6.70 & 6.63 & 6.64 &  \\ 
P323+12 & 6.61 & 6.58& 6.59 & 6.58 & 6.59 & 6.58\\ 
J0109-3047 & 6.77 & 6.75& 6.76 & 6.69 &   & 6.74\\ 
    \end{tabular}
    \caption{BEL-based redshifts for all our objects. SDSS quasars are designated by their plate-mjd-fiber identification to make their search easier in the SDSS database. The full table is published as an online supplementary material. \label{tab:all_redshifts}}
    
\end{table}

\section{Catastrophic redshift errors between our method and DR12Q tabulated values}

\begin{figure*}
    \centering
    \includegraphics[width =0.45\textwidth]{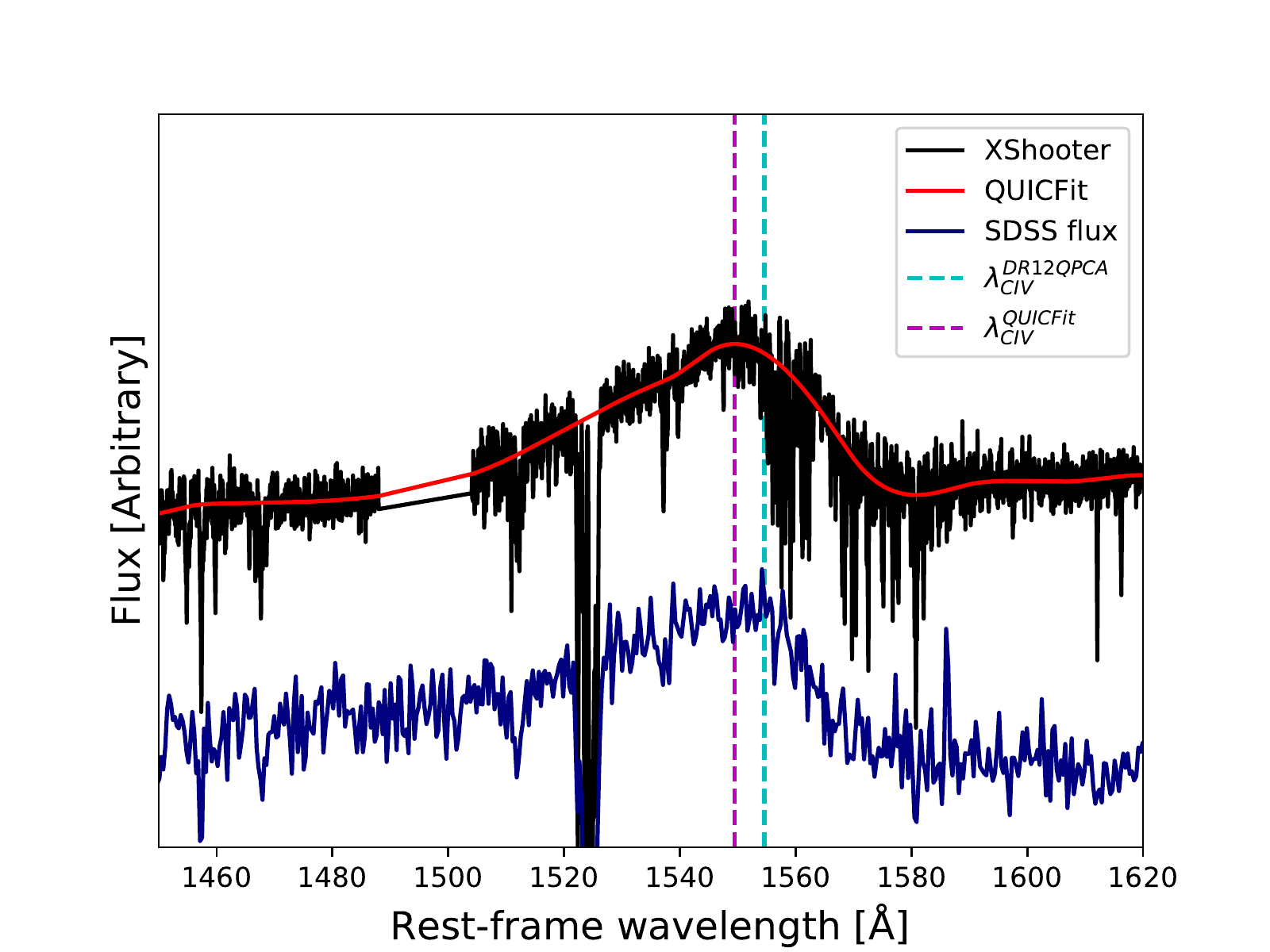}
    \includegraphics[width=0.45\textwidth]{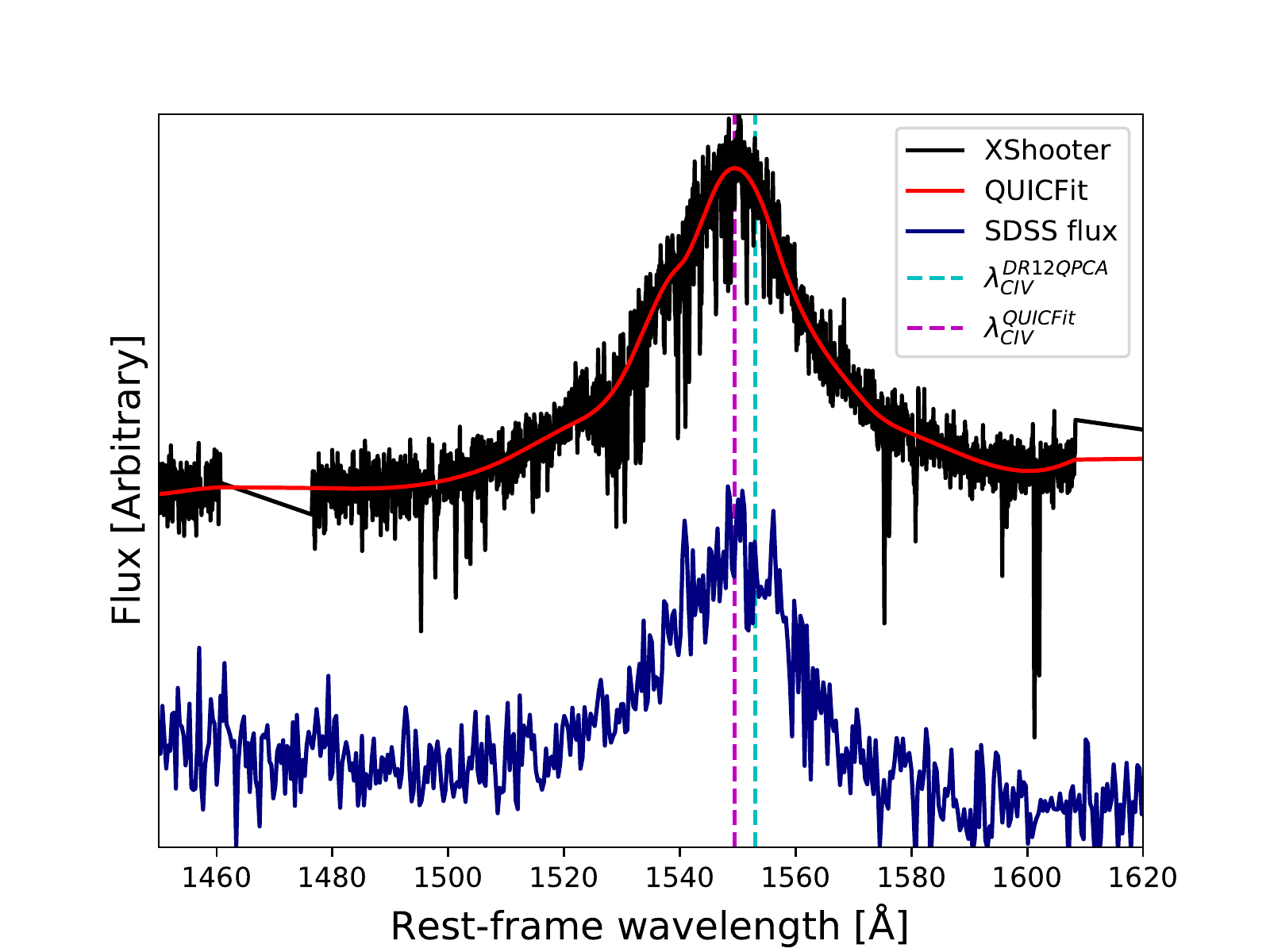}
    \\
    \includegraphics[width=0.45\textwidth]{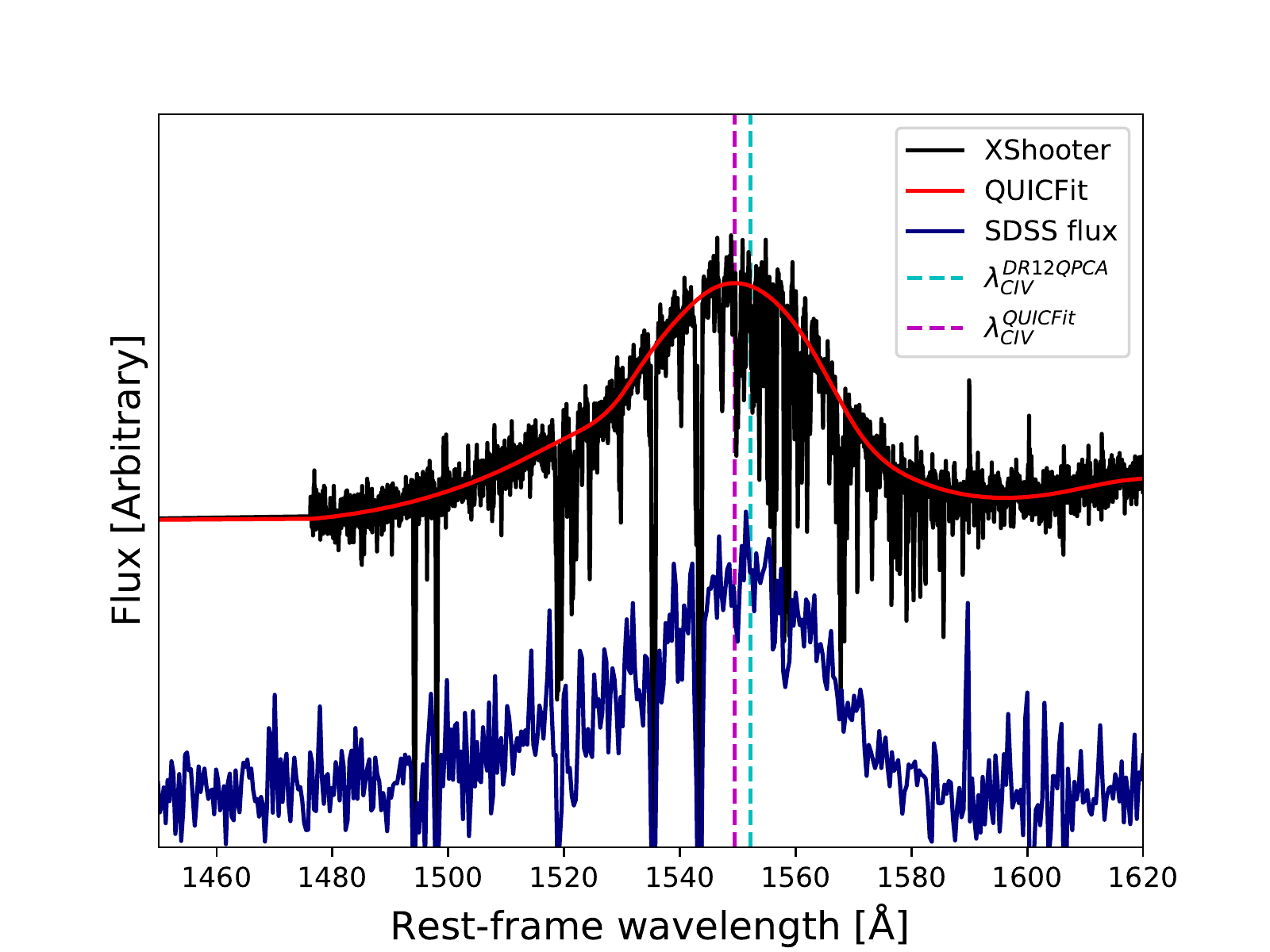}
    \includegraphics[width=0.45\textwidth]{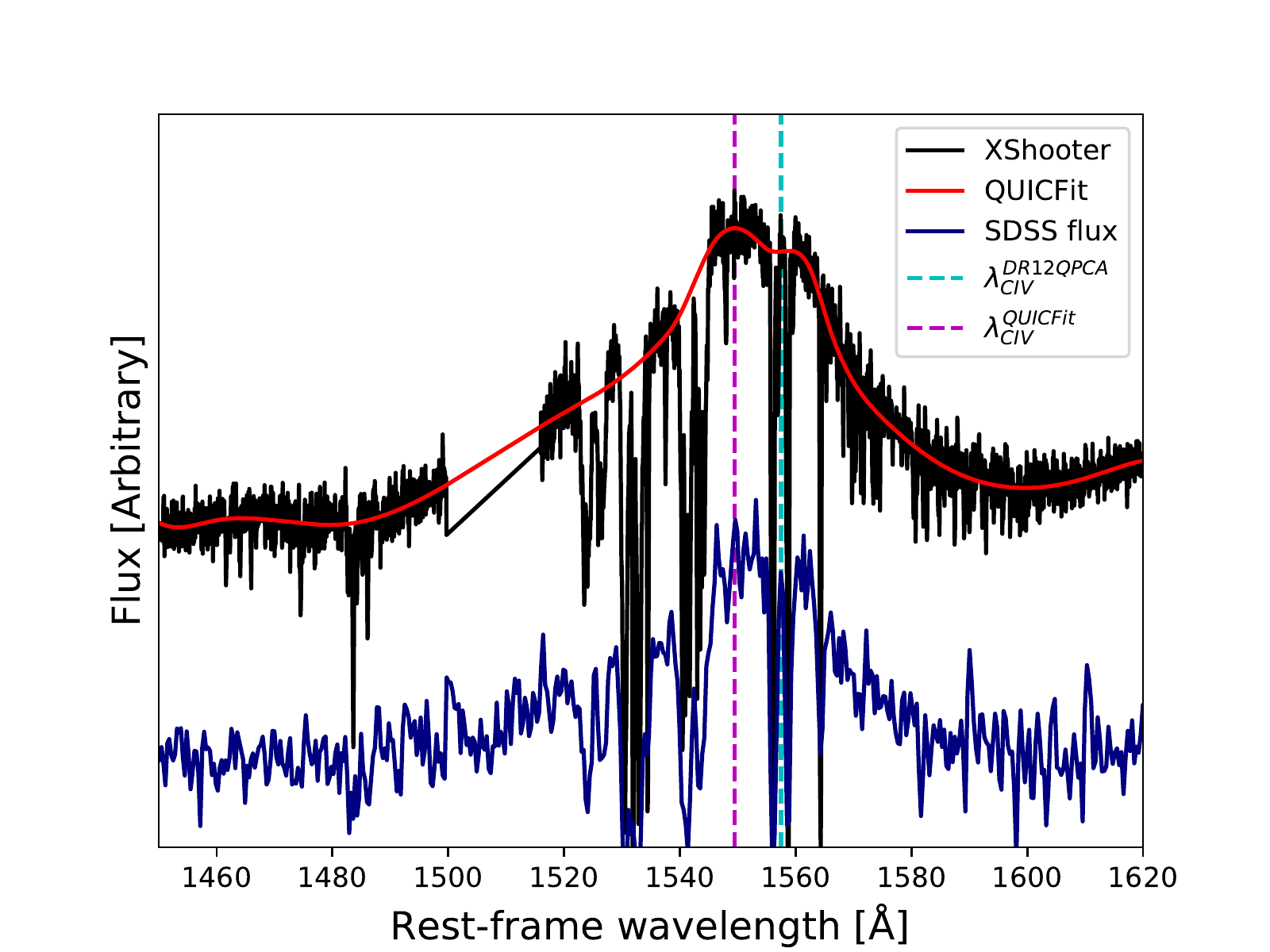}
    \\ 
    \includegraphics[width=0.45\textwidth]{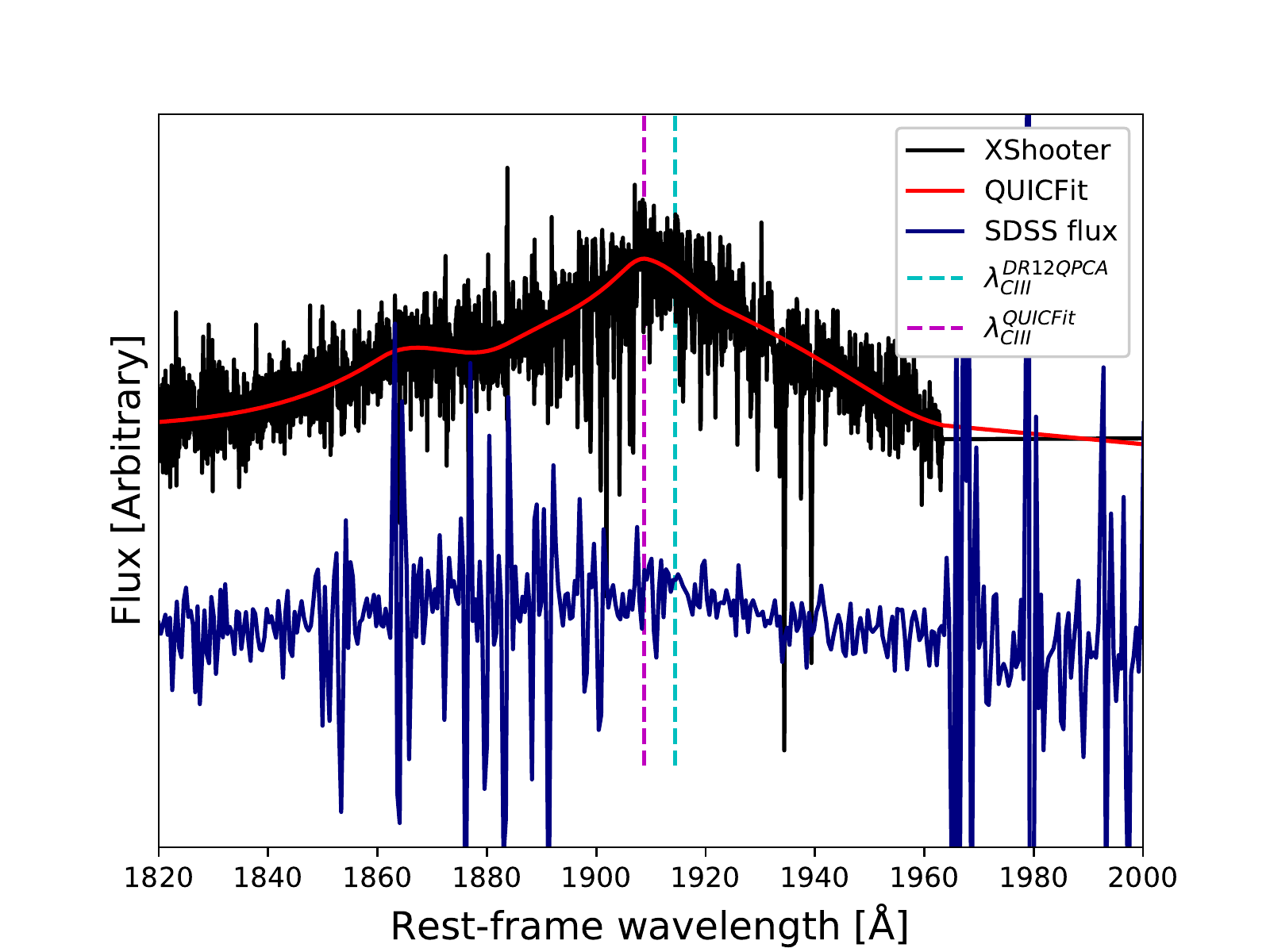}
    \includegraphics[width=0.45\textwidth]{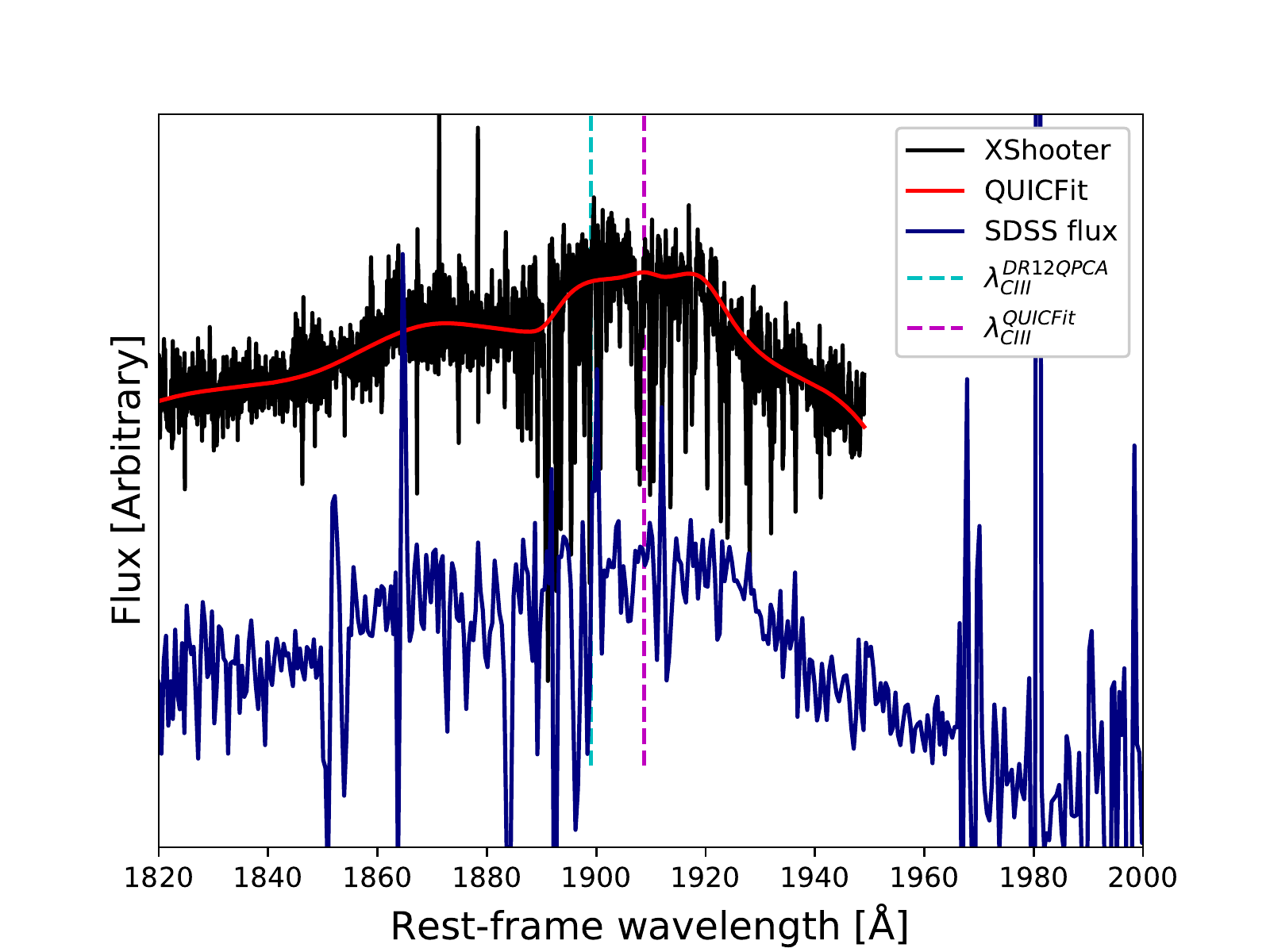} 
        \includegraphics[width=0.45\textwidth]{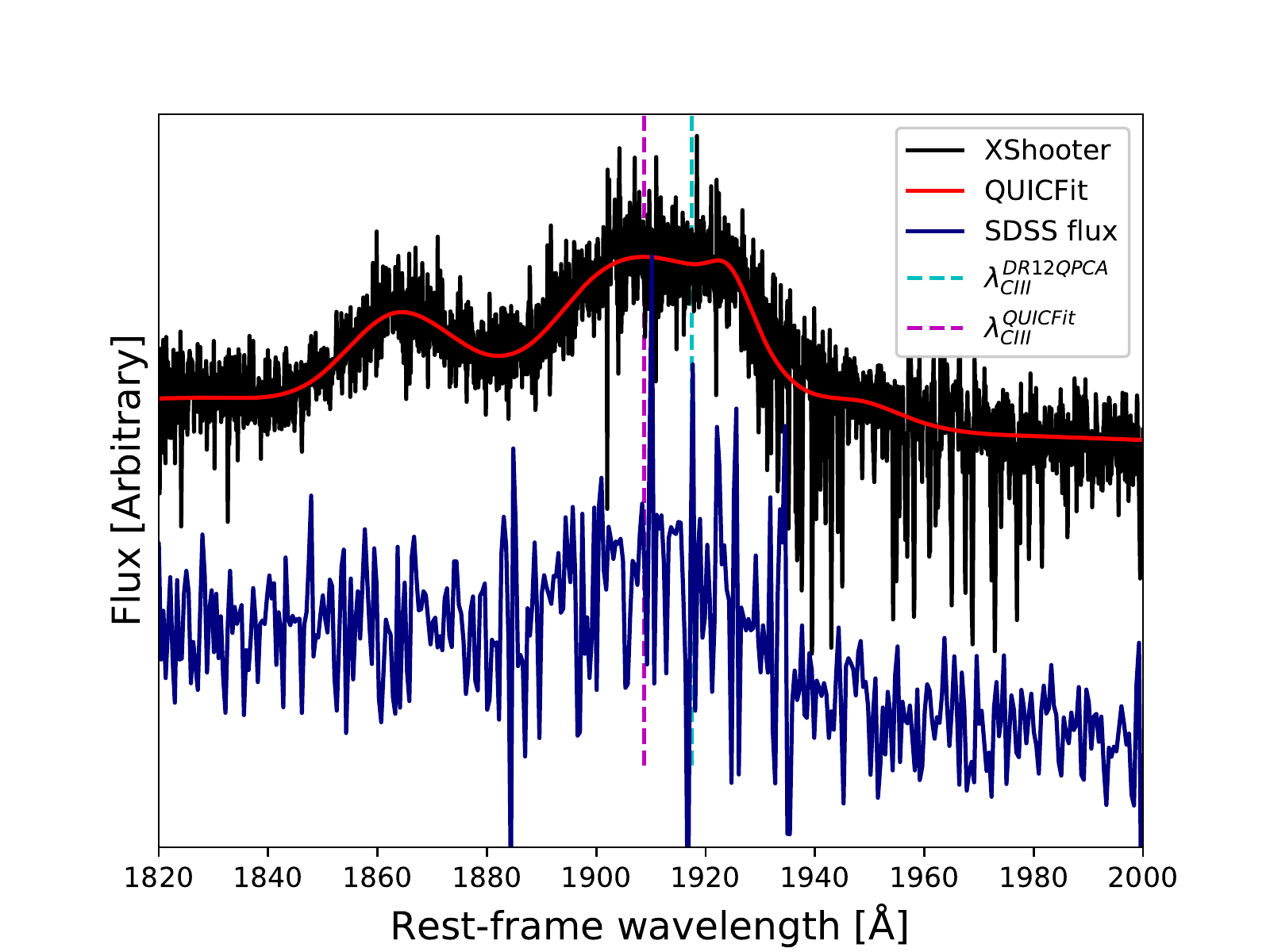}
    \includegraphics[width=0.45\textwidth]{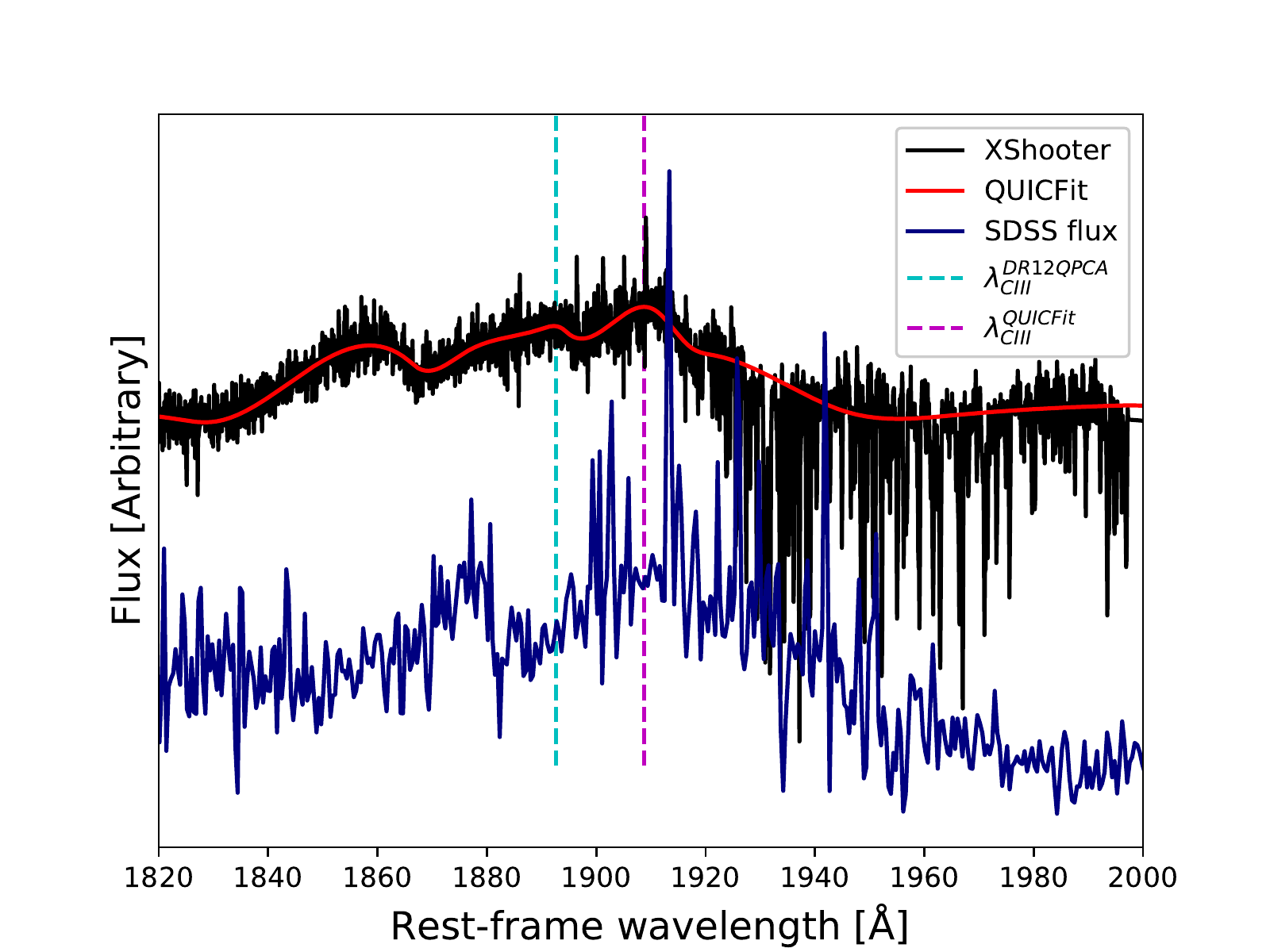} 
    \caption{Quasars in the XQ100-SDSS overlap for which the difference between the DR12Q PCA-based redshift and our solution for the \cfour \, or the \cthree \, line exceeds $500$ km s $^{-1}$ (see Fig. \ref{fig:comparison_SDSS_XQ100}). }
    \label{fig:catastrophic_SDSS_XQ100_CIV}
\end{figure*}

\begin{figure*}
    \centering
    \includegraphics[width=0.45\textwidth]{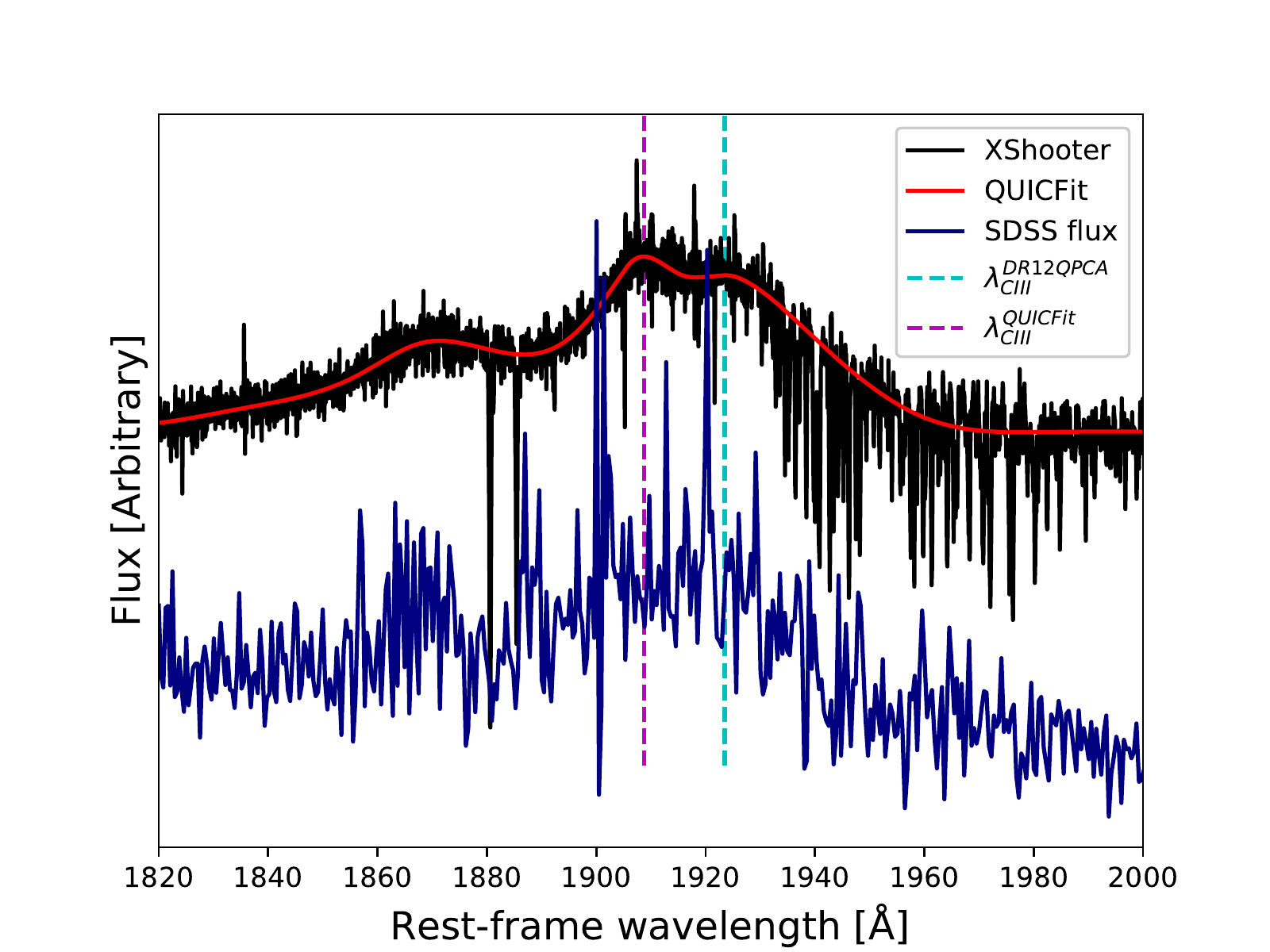}
    \includegraphics[width=0.45\textwidth]{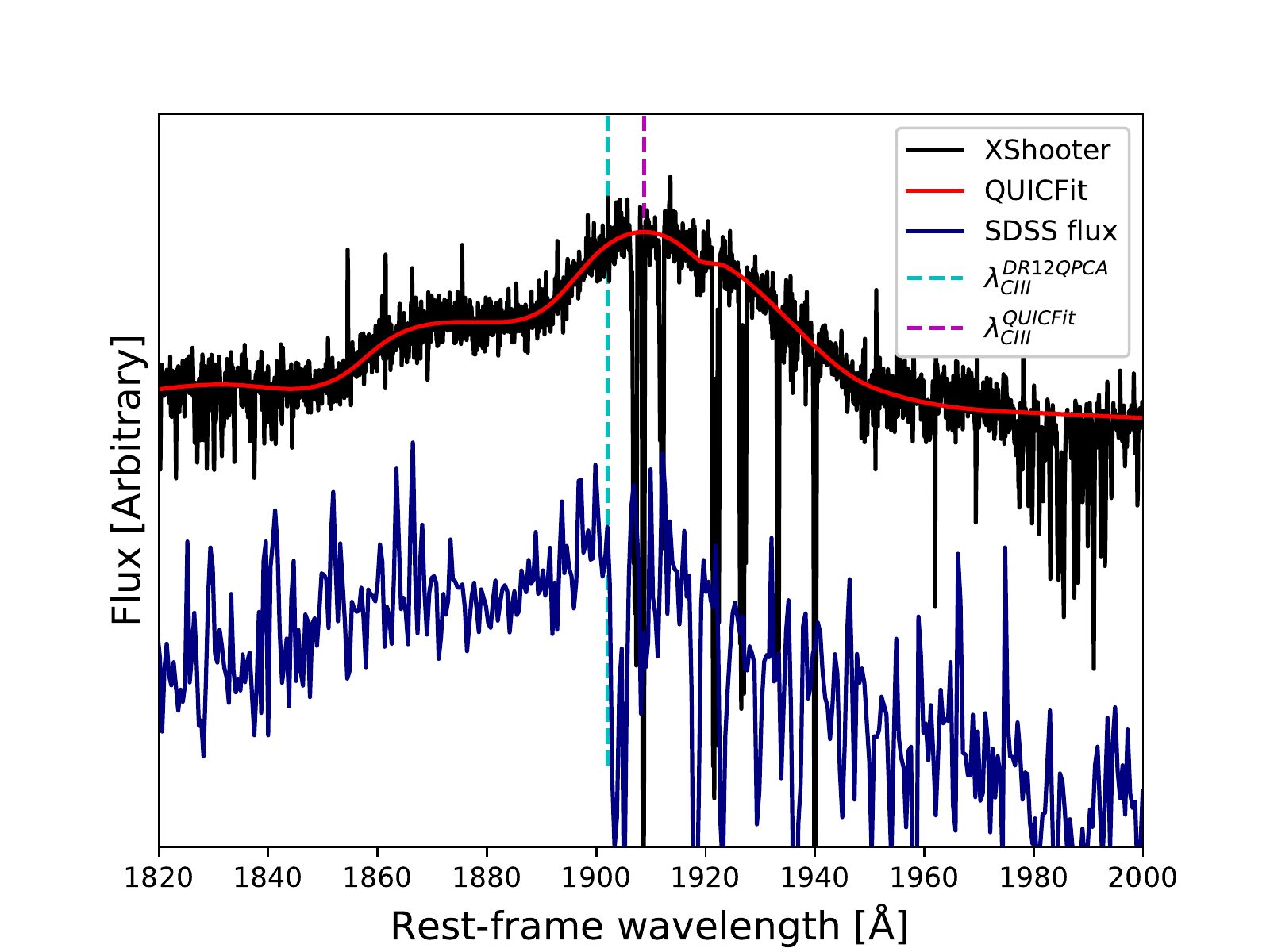} \\
    \includegraphics[width=0.45\textwidth]{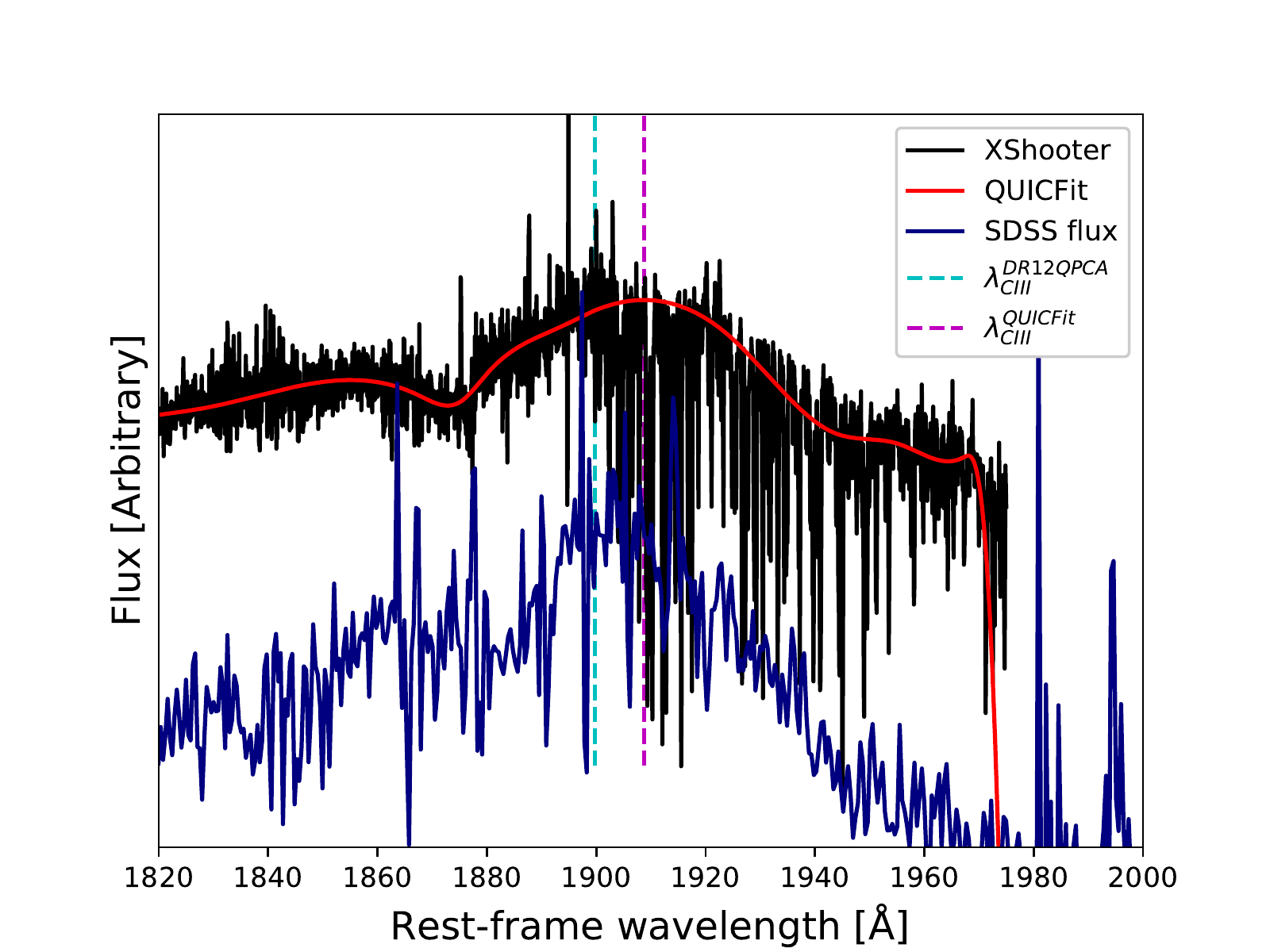}
    \includegraphics[width=0.45\textwidth]{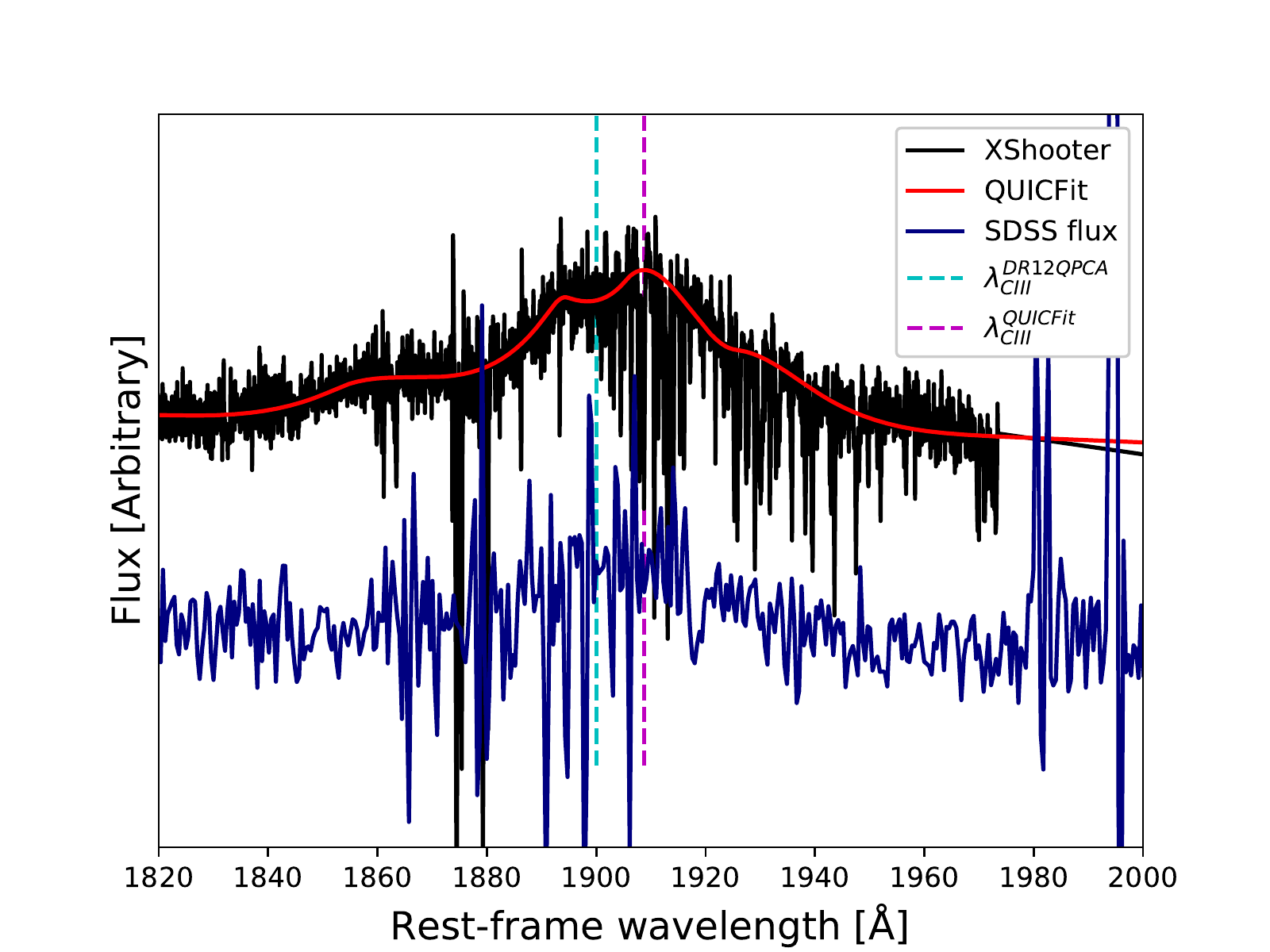} \\
    \includegraphics[width=0.45\textwidth]{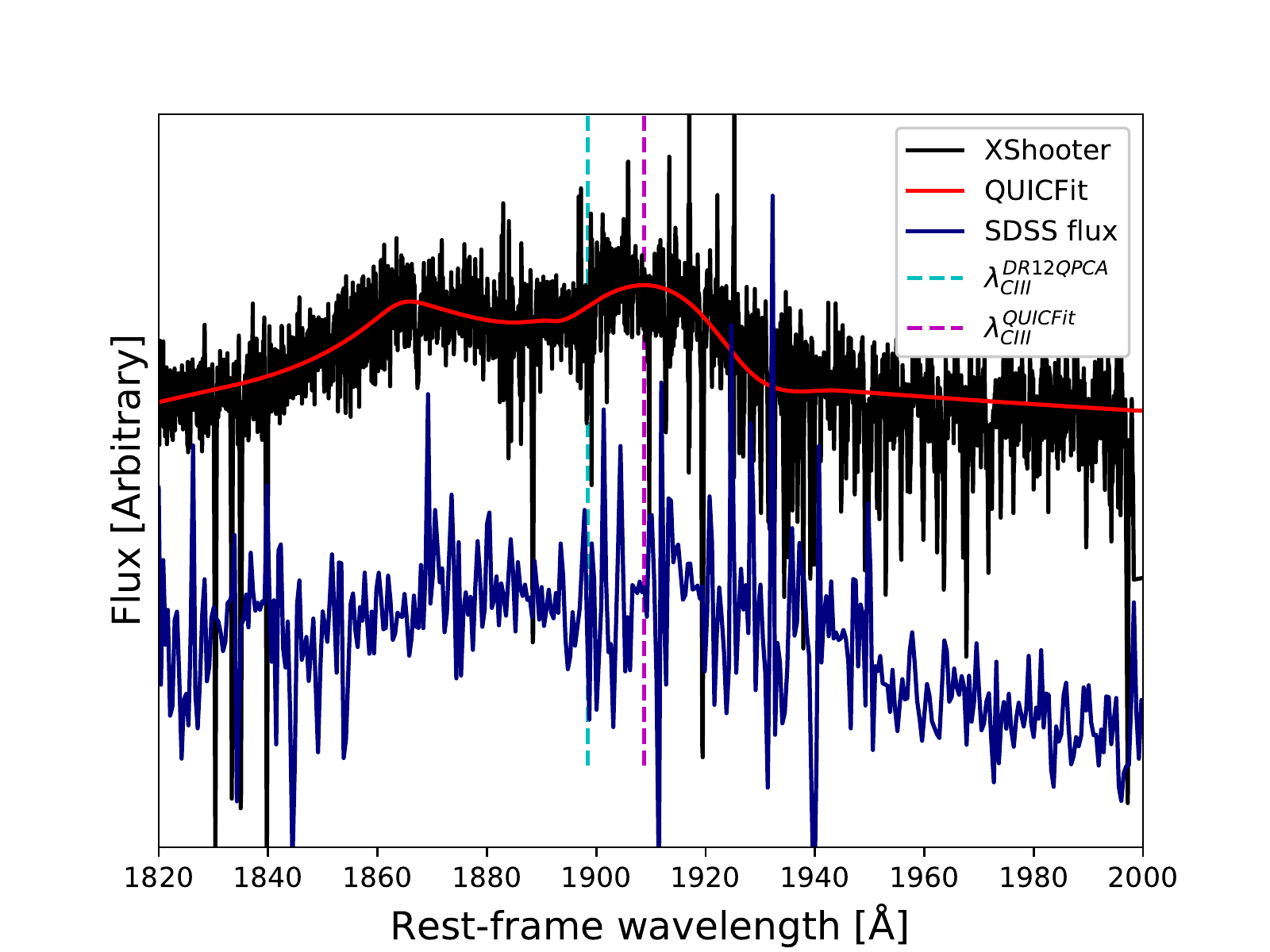}
    \includegraphics[width=0.45\textwidth]{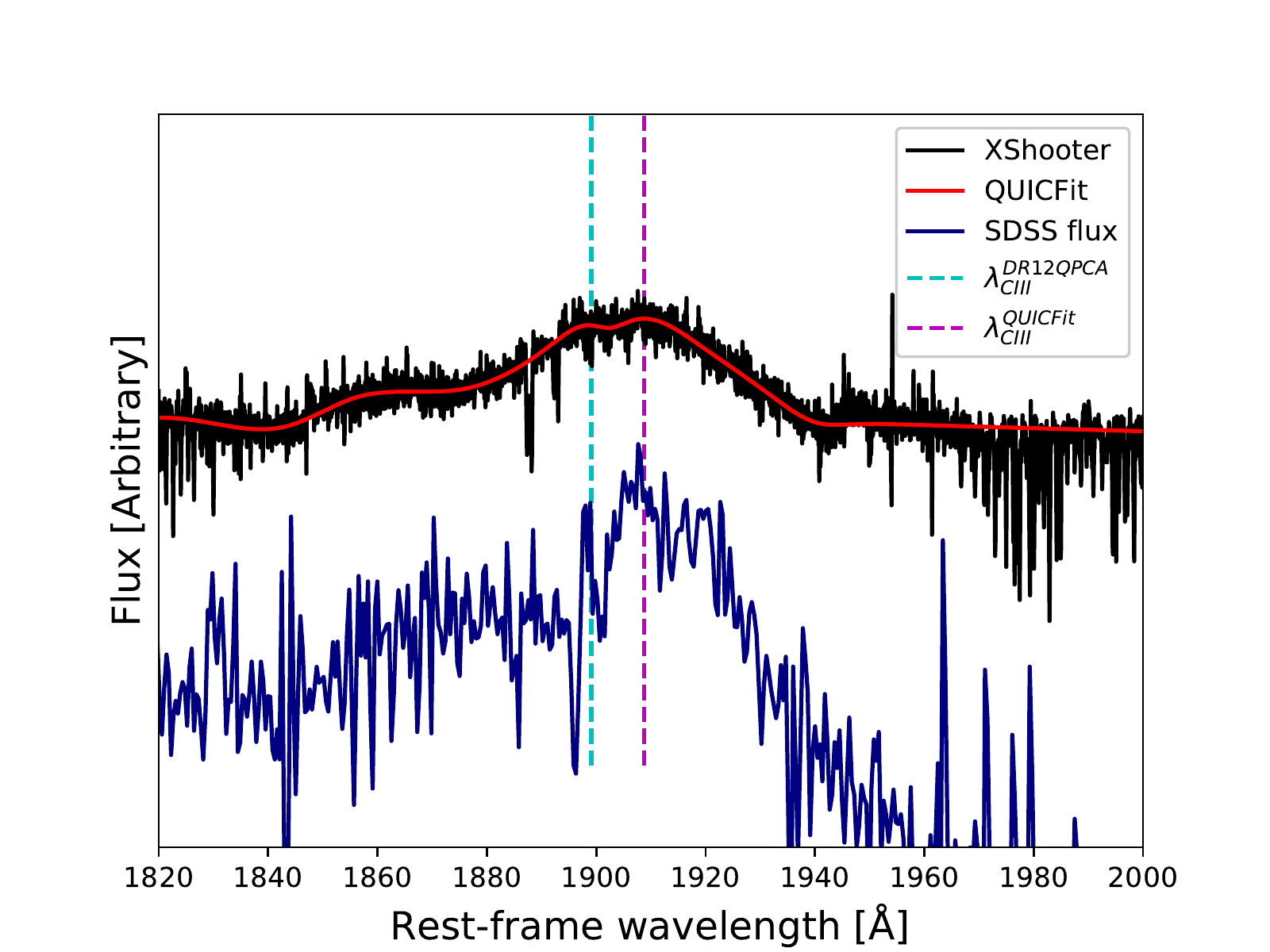} \\ 
    \includegraphics[width=0.45\textwidth]{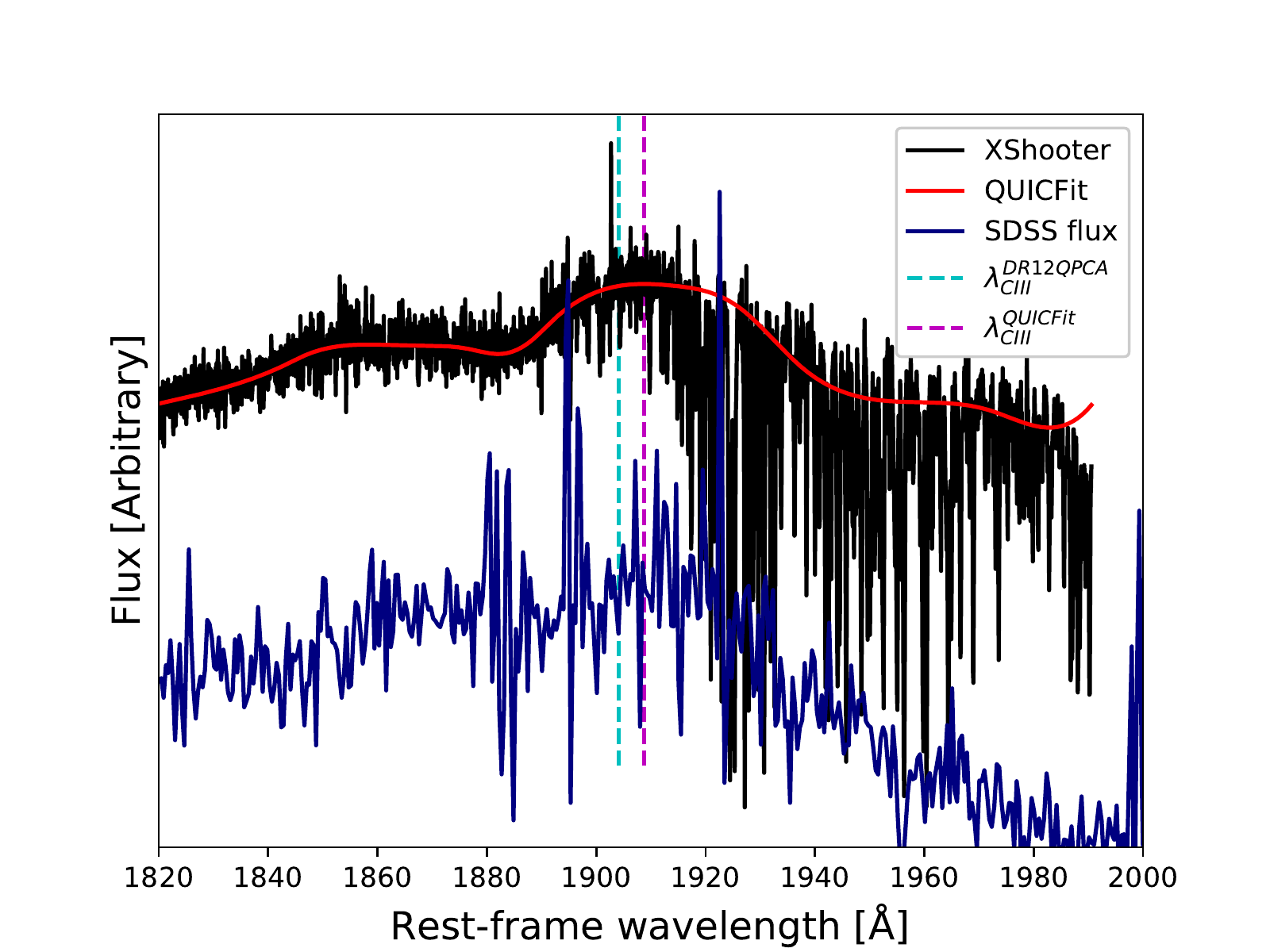}\\
    \caption{Quasars in the XQ100-SDSS overlap for which the difference between the DR12Q PCA-based redshift and our solution for the \cthree line exceeds $500$ km s $^{-1}$ (see Fig. \ref{fig:comparison_SDSS_XQ100}).  }
        \label{fig:catastrophic_SDSS_XQ100_CIII}
\end{figure*}

\section{Relative velocity shifts including \sifour\, and \cthree\,}
\begin{figure*}
    \centering
    \includegraphics[width=\textwidth]{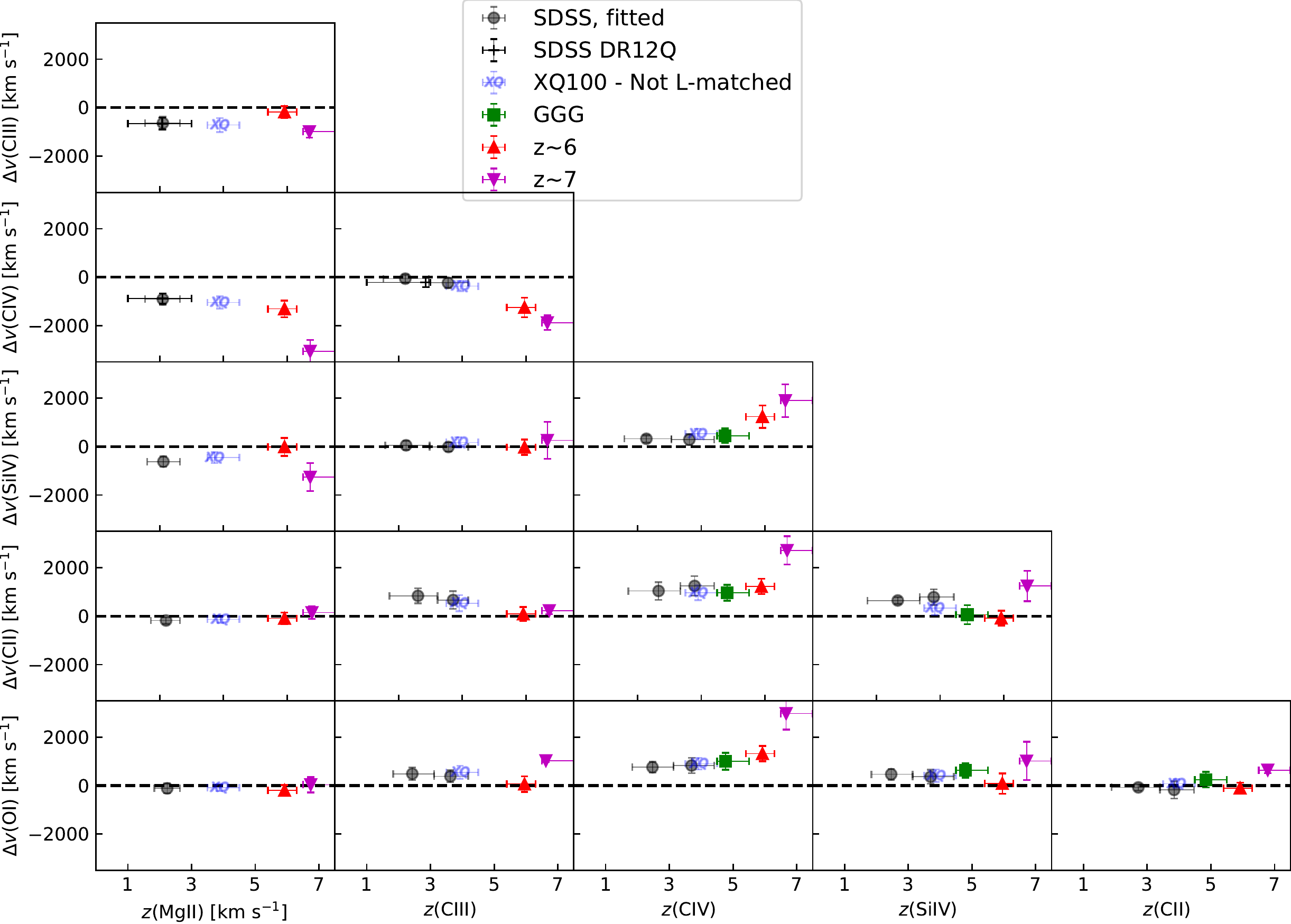}
    \caption{Velocity shifts of rest-frame UV BELs across redshift. The errors are computed by bootstrapping using samples size of the z6 sample to make the errors comparable.  \label{fig:evol_all} }
\end{figure*}

We report on Fig. \ref{fig:evol_all} the relative velocity shifts of all lines, including the \sifour\, and \cthree\,  complexes. It must be noted that, because shifts are derived from the peak of the complex, the shift of these two complexes are mostly driven by the relative amplitude and thus abundance of the underlying emission lines evolving with redshift.

\bsp	
\label{lastpage}
\end{document}